\documentclass{aa}
\usepackage{graphics,psfig}

\begin{document}      

   \title{Ram-pressure stripped molecular gas in the Virgo spiral galaxy NGC~4522\thanks{Based on IRAM 30m HERA observations}}

   \author{B.~Vollmer\inst{1}, J.~Braine\inst{2}, C.~Pappalardo\inst{1}, P.~Hily-Blant\inst{3}}

   \offprints{B.~Vollmer, e-mail: bvollmer@astro.u-strasbg.fr}

   \institute{CDS, Observatoire astronomique de Strasbourg, 11, rue de l'universit\'e,
	      67000 Strasbourg, France \and
	      Laboratoire d'Astrophysique de Bordeaux (OASU), Universit\'e de Bordeaux, UMR 5804, CNRS/INSU, B.P. 89, F-33270 Floirac, France \and
	      IRAM, Domaine Universitaire, 300 rue de la Piscine, 38406 Saint-Martin d'H\`eres, France
              }

   \date{Received / Accepted}

   \authorrunning{Vollmer et al.}
   \titlerunning{Stripped molecular gas in NGC~4522}

\abstract{
IRAM 30m $^{12}$CO(1--0) and $^{12}$CO(2--1) HERA observations are presented for the ram-pressure stripped Virgo spiral
galaxy NGC~4522. The CO emission is detected in the galactic disk and the extraplanar
gas. The extraplanar CO emission follows the morphology of the atomic gas closely but is less extended.
The CO maxima do not appear to correspond to regions where there is peak massive star formation as probed by
H$\alpha$ emission.
The presence of molecular gas is a necessary but not sufficient condition
for star formation. Compared to the disk gas, the molecular fraction of the extraplanar gas is 30\,\% lower
and the star formation efficiency of the extraplanar gas is about 3 times lower.
The comparison with an existing dynamical model extended by a recipe for
distinguishing between atomic and molecular gas shows that a significant part of the gas is stripped
in the form of overdense arm-like structures.
It is argued that the molecular fraction depends on the square root of the total large-scale density. 
Based on the combination of the CO/H$\alpha$ and an analytical model, the total gas density is estimated to
be about 4 times lower than that of the galactic disk.
Molecules and stars form within this dense gas according to the same laws as in the galactic disk, 
i.e. they mainly depend on the total large-scale gas density.
Star formation proceeds where the local large-scale gas density is highest. Given the complex
3D morphology this does not correspond to the peaks in the surface density.
In the absence of a confining gravitational potential, the stripped gas
arms will most probably disperse; i.e. the density of the gas will decrease and star formation will cease.
\keywords{
Galaxies: individual: NGC~4522 -- Galaxies: interactions -- Galaxies: ISM
-- Galaxies: kinematics and dynamics -- Stars: formation -- Radio lines: ISM
}
}

\maketitle

\section{Introduction \label{sec:intro}}

The best-studied case of active ram-pressure stripping of a cluster galaxy is NGC~4522
located in the Virgo cluster. Only in this cluster is the resolution of radio telescopes
sufficient ($\sim 20'' = 1.6$~kpc\footnote{We use a distance of 17~Mpc for the Virgo cluster.})
for a detailed analysis of the gas morphology and kinematics.
NGC~4522 is a rather small ($D_{25}=4'=20$~kpc) edge-on Sc galaxy with a rotation velocity of
$\sim 100$~km\,s$^{-1}$. It is strongly H{\sc i} deficient ($DEF=0.6$; Helou et al. 1984). 
Its projected distance to the
cluster center (M~87) is large ($\sim 1$~Mpc), and its radial velocity with respect to the
Virgo cluster mean is high ($1150$~km\,s$^{-1}$). H{\sc i} and H$\alpha$ observations (Kenney et al. 2004; 
Kenney \& Koopmann 1999) show a
heavily truncated gas disk at a radius of 3~kpc, which is $\sim 40$\% of the optical radius,
and a significant amount of extraplanar gas to the west 
of the galactic disk. The one-sided extraplanar atomic gas distribution shows high column
densities, comparable to those of the adjacent galactic disk.
The 6~cm polarized radio continuum emission shows a maximum at the eastern edge of the galactic disk,
on the opposite side of the extraplanar gas and star formation.
Since the stellar disk is symmetric and undisturbed (Kenney \& Koopmann 1999), 
a tidal interaction is excluded as the origin of the peculiar 
gas distribution of NGC~4522. 
Thus,  this galaxy undergoes ram-pressure 
stripping due to the galaxy's rapid motion within the hot and tenuous intracluster gas (ICM) of the 
Virgo cluster.

Vollmer et al. (2006) made a dynamical model that includes the effects of ram pressure for NGC~4522.
The model successfully reproduces the large-scale gas distribution and the velocity field.
By assuming a Gaussian distribution of relativistic electrons, they obtained the distribution of polarized radio 
continuum emission, which reproduces the VLA observations of polarized radio continuum emission at 6~cm. 
The observed maximum of the polarized radio continuum emission is successfully reproduced.
The eastern ridge of polarized radio continuum emission is therefore due to ram pressure compression of the 
interstellar medium (ISM) and its magnetic field.  
The dynamical model and the analysis of the stellar populations of the outer gas-free disk
using optical spectra and UV photometry (Crowl \& Kenney 2006)
indicate that the ram pressure maximum occurred only $\sim 50$--$100$~Myr ago.
This scenario has one important caveat: the large projected distance of NGC~4522 ($1$~Mpc)
to the center of the Virgo cluster (M87). Assuming a static smooth ICM and standard values
for the ICM density and the galaxy velocity, the ram pressure at that location
seems to be too low by an order of magnitude to produce the observed truncation of the gas disk. 
A natural explanation for the enhanced ram pressure efficiency  is that 
the intracluster medium is not static but moving due to the infall of the M49
group of galaxies from behind (Kenney et al. 2004, Vollmer et al. 2004, Vollmer et al. 2006). 
In this case the galaxy has just passed the region of highest intracluster medium velocity.

While we know from the H$\alpha$ observations (Kenney et al. 2004, Kenney \& Koopmann 1999)
that stars are forming in the extraplanar gas, we do not know the distribution of molecular gas
in these regions. How does the complex multiphase interstellar medium respond to ram pressure
stripping? Can we model the molecular gas content during the interaction using simplified recipes?
Can dense molecular gas decouple from the ram pressure wind as suggested for NGC~4438
(Vollmer et al. 2005)? 
In this article we present IRAM 30m $^{12}$CO(1--0) and $^{12}$CO(2--1) HERA observations of NGC~4522
to investigate the fate of the stripped gas.

We present our CO observations in Sec.~\ref{sec:observations} followed by the observational 
results in Sec.~\ref{sec:results}. The detection of ram pressure wind decoupled molecular
clouds is reported in Sec.~\ref{sec:decoupled}. In Sec.~\ref{sec:comparison} we compare our CO observations to
existing H{\sc i} and H$\alpha$ emission distributions (Kenney et al. 2004) and to the dynamical
model of Vollmer et al. (2006). The molecular fraction and star formation efficiencies are
discussed in Sec.~\ref{sec:discussion} and we give our conclusions in Sec.~\ref{sec:conclusions}.

\section{Observations \label{sec:observations}}

\begin{figure}
	\resizebox{\hsize}{!}{\includegraphics{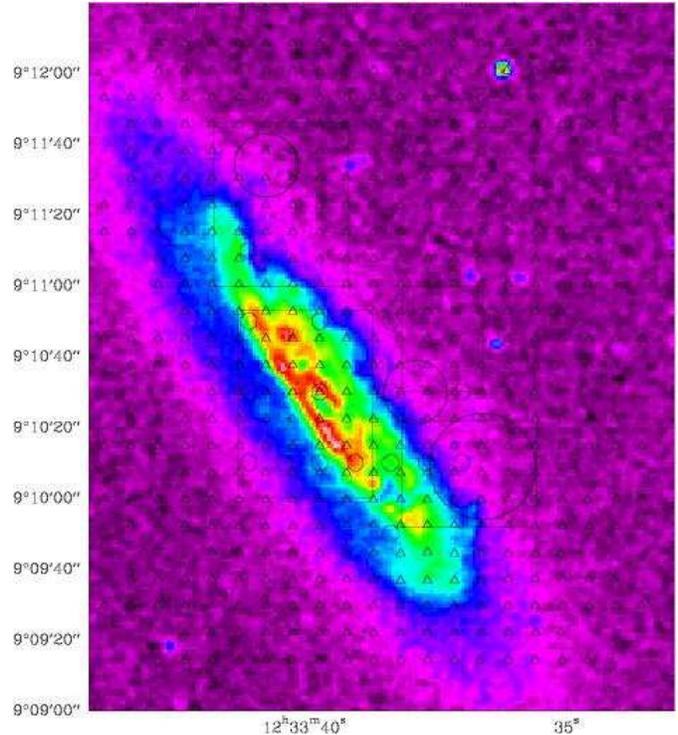}}
	\caption{IRAM 30m pointings on a DSS B band image. Triangles: $^{12}$CO(2-1) HERA observations. 
	  Small circles: $^{12}$CO(1-0) observations. 
	  Large circles: Binned HERA pointings (see Fig.~\ref{fig:co2}).
	  The large boxes correspond to Fig.~\ref{fig:cospec1} to Fig.~\ref{fig:cospec4}.
	} \label{fig:pointingdss}
\end{figure}

\begin{figure}
	\resizebox{\hsize}{!}{\includegraphics{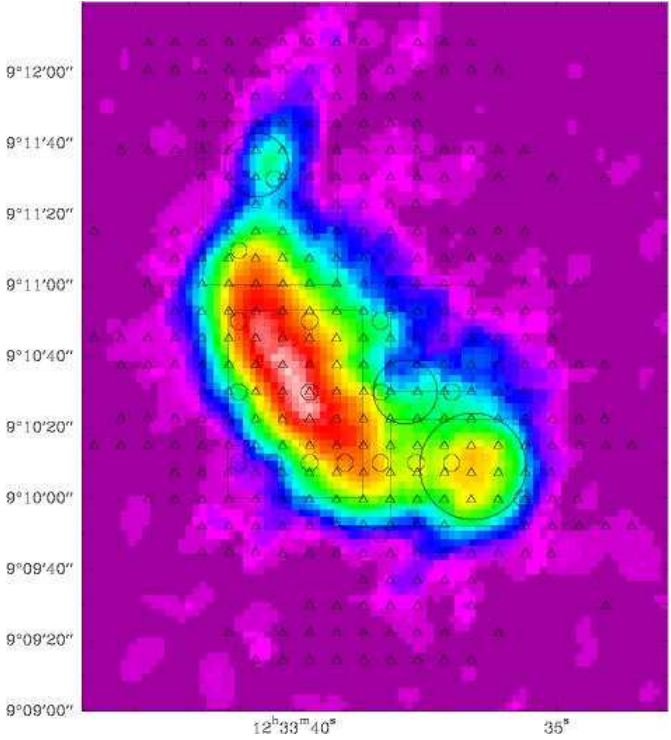}}
	\caption{$^{12}$CO(2-1) HERA pointings (see Fig.~\ref{fig:pointingdss}) with low noise levels on the H{\sc i}
	  emission distribution (from Kenney et al. 2004). 
	  The large boxes correspond to Fig.~\ref{fig:cospec1} to Fig.~\ref{fig:cospec4}.
	} \label{fig:pointinghi}
\end{figure}

The observations of the CO(1--0) and CO(2--1) lines, with rest frequencies of 115.271204
and 230.53799 GHz respectively, were carried out at the 30~meter  millimeter-wave
telescope on Pico Veleta (Spain) run by the Institut de RadioAstronomie
Millim\'etrique (IRAM).  The CO(2--1) observations used the HERA  multi-beam
array, with 3 $\times$ 3 dual-polarization receivers, and the WILMA  autocorrelator backend
with 2MHz spectral resolution.  The CO(1--0) observations used the  single-pixel "AB" receivers
and the 1MHz filterbanks as backends.  The spectral resolution is 2.6~km\,s$^{-1}$ in
both cases.  The HERA observations were made in February and March  2006 and the
CO(1--0) in November 2006.
In both cases, a nutating secondary ("wobbler") was used with a throw  of 180--200 arcseconds
in order to be clear of any emission from the galaxy.
The positions observed in each line are indicated in Figs.~\ref{fig:pointingdss}
and  \ref{fig:pointinghi} as  triangles for CO(2--1) and circles for CO(1--0).

Data reduction was straightforward, eliminating any obviously bad channels
and excluding the spectra taken under particularly poor conditions (system temperature
over 1000~K).  Spectra were then summed position by position.
System temperatures of the final spectra ranged from 200 to 500~K on the Ta$^*$ scale.
All spectra are presented on the main beam temperature scale,  assuming telescope 
main-beam and forward efficiencies of
$\eta_{\rm mb}=0.54$ and $\eta_{\rm for}=0.90$ for HERA and
$\eta_{\rm mb}=0.74$ and $\eta_{\rm for}=0.95$ for the CO(1--0) line.
The spectra near map edges with noise levels greater than 28 mK (T$_{mb}$ scale) are left out of 
Fig.~\ref{fig:pointinghi}.
At the assumed distance of NGC~4522, 17~Mpc, the CO(2--1) and CO (1--0) beams correspond to
0.9 and 1.7~kpc respectively.  In order to convert CO integrated intensities into molecular gas masses,
we have assumed a $N({\rm H}_2) / I_{\rm CO(2-1)}$ ratio 
of $2 \times 10^{20}$~H$_2$ mol~cm$^{-2}$ per K kms$^{-1}$.
Our conclusions, however, do not depend strongly on the $N({\rm H}_2) / I_{\rm CO}$ ratio within
reasonable variations.

\section{Disk and extraplanar molecular gas \label{sec:results}}

\subsection{CO spectra}

\begin{figure}
	\resizebox{\hsize}{!}{\includegraphics{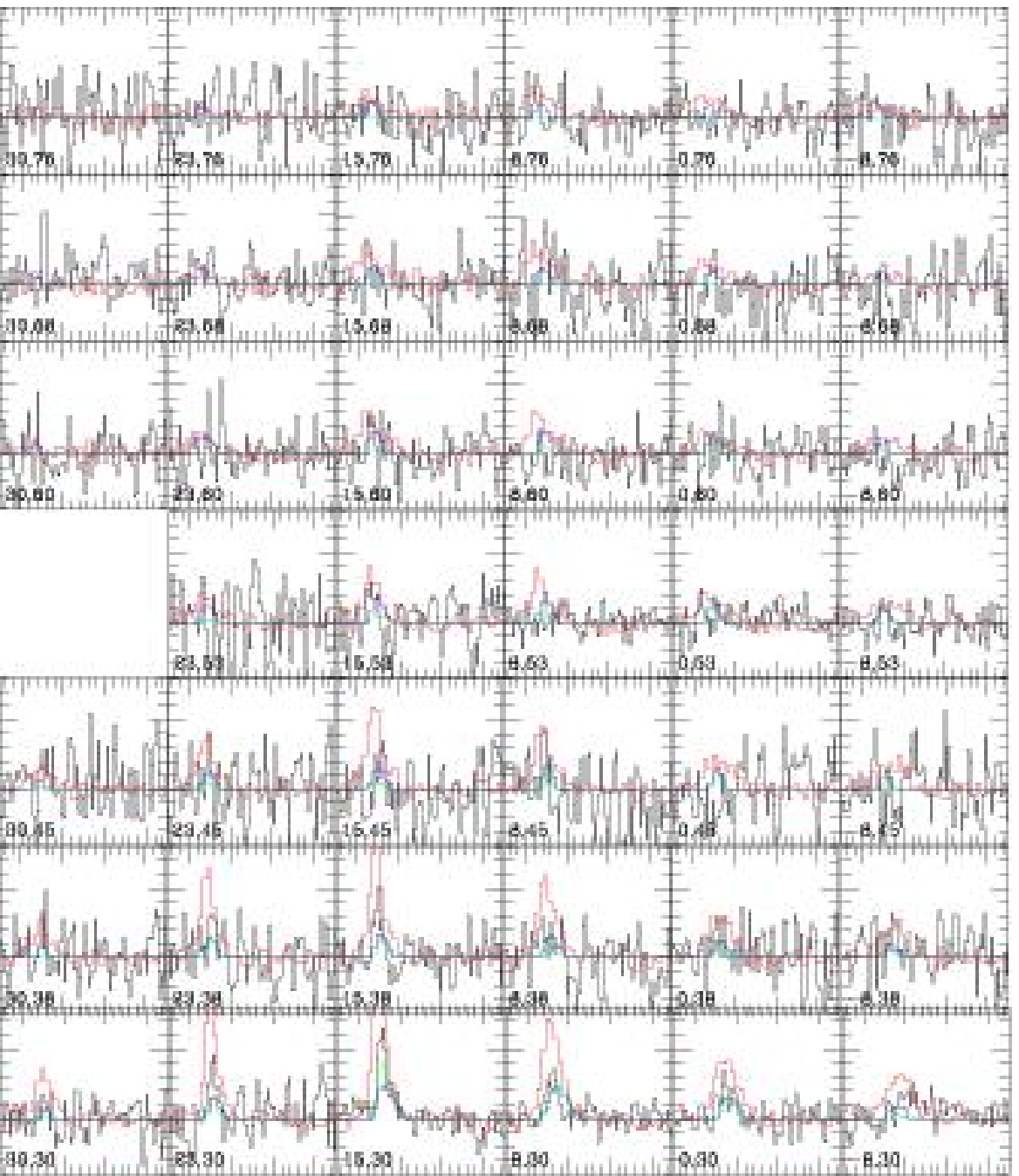}}
	\caption{IRAM 30m HERA $^{12}$CO(2-1) spectra of the northeastern box of Fig.~\ref{fig:pointingdss}
	  (black lines). H{\sc i} spectra from Kenney et al. (2004) (red lines). 
	  CO (green lines) and H{\sc i} (blue lines) spectra from the dynamical model of Vollmer et al. (2006).
	  The velocity scale is from 2150 to 2550~km\,s$^{-1}$ and the main-beam temperature scale 
	  from -40~mK to 80 mK (T$_{mb}$ scale).
	} \label{fig:cospec1}
\end{figure}

\begin{figure}
	\resizebox{\hsize}{!}{\includegraphics{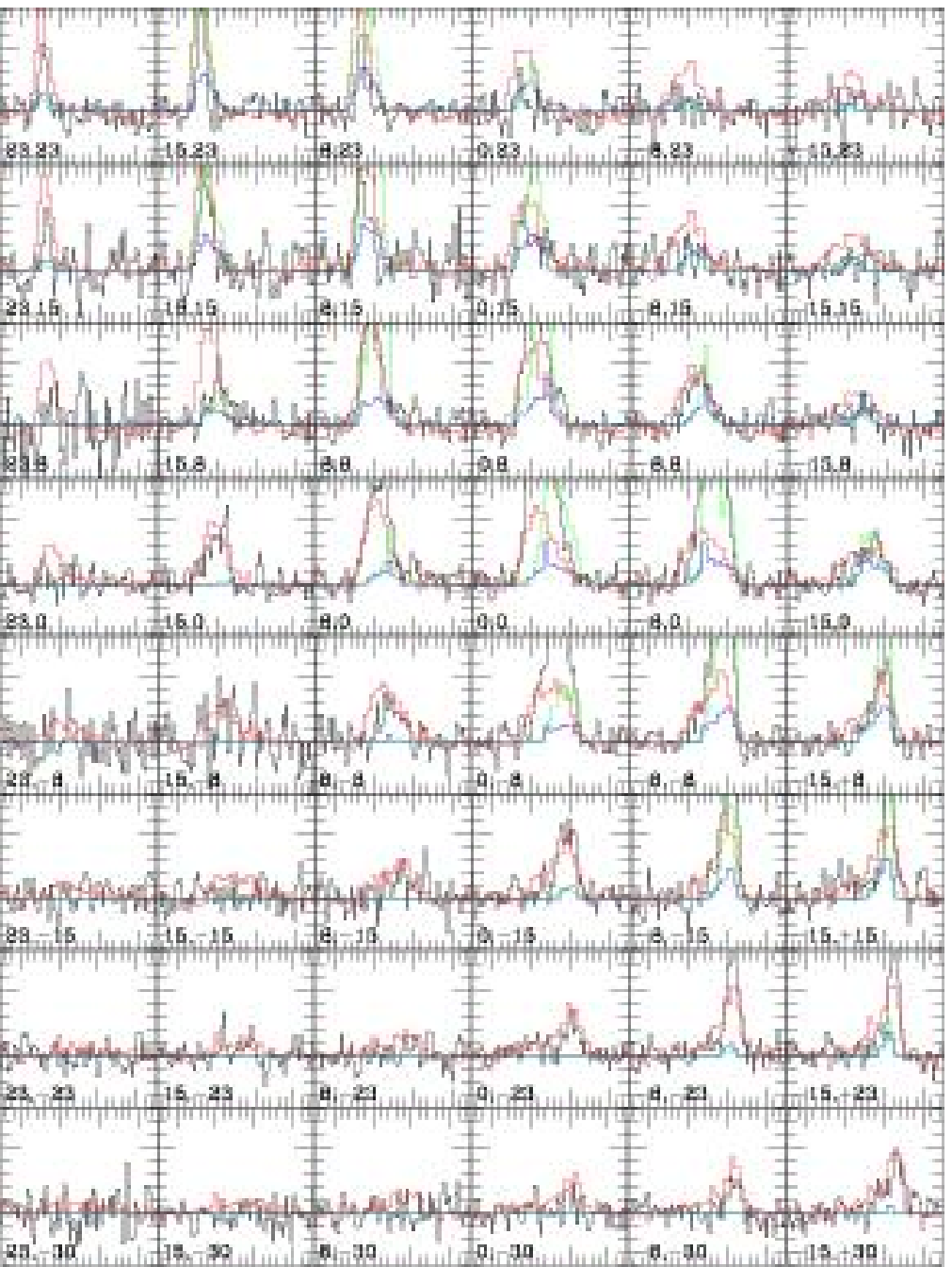}}
	\caption{IRAM 30m HERA $^{12}$CO(2-1) spectra of the southeastern box of Fig.~\ref{fig:pointingdss}
	  (black lines). H{\sc i} spectra from Kenney et al. (2004) (red lines). 
	  CO (green lines) and H{\sc i} (blue lines) spectra from the dynamical model of Vollmer et al. (2006).
	  The velocity scale is from 2150 to 2550~km\,s$^{-1}$ and the main-beam temperature scale 
	  from -40~mK to 80 mK (T$_{mb}$ scale).
	} \label{fig:cospec2}
\end{figure}

\begin{figure}
	\resizebox{\hsize}{!}{\includegraphics{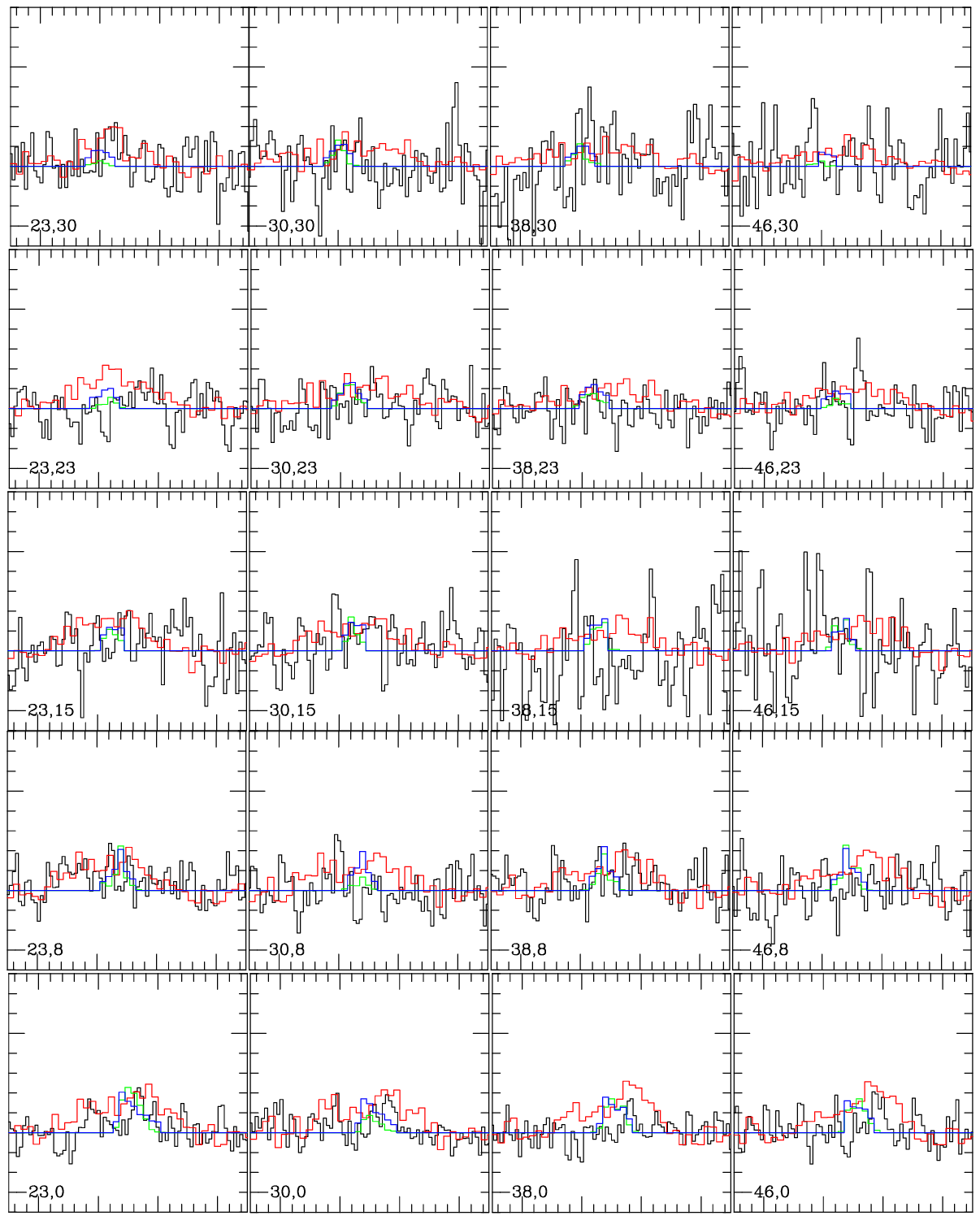}}
	\caption{IRAM 30m HERA $^{12}$CO(2-1) spectra of the northwestern box of Fig.~\ref{fig:pointingdss}
	  (black lines). H{\sc i} spectra from Kenney et al. (2004) (red lines). 
	  CO (green lines) and H{\sc i} (blue lines) spectra from the dynamical model of Vollmer et al. (2006).
	  The velocity scale is from 2150 to 2550~km\,s$^{-1}$ and the main-beam temperature scale 
	  from -40~mK to 80 mK (T$_{mb}$ scale).
	} \label{fig:cospec3}
\end{figure}

\begin{figure}
	\resizebox{\hsize}{!}{\includegraphics{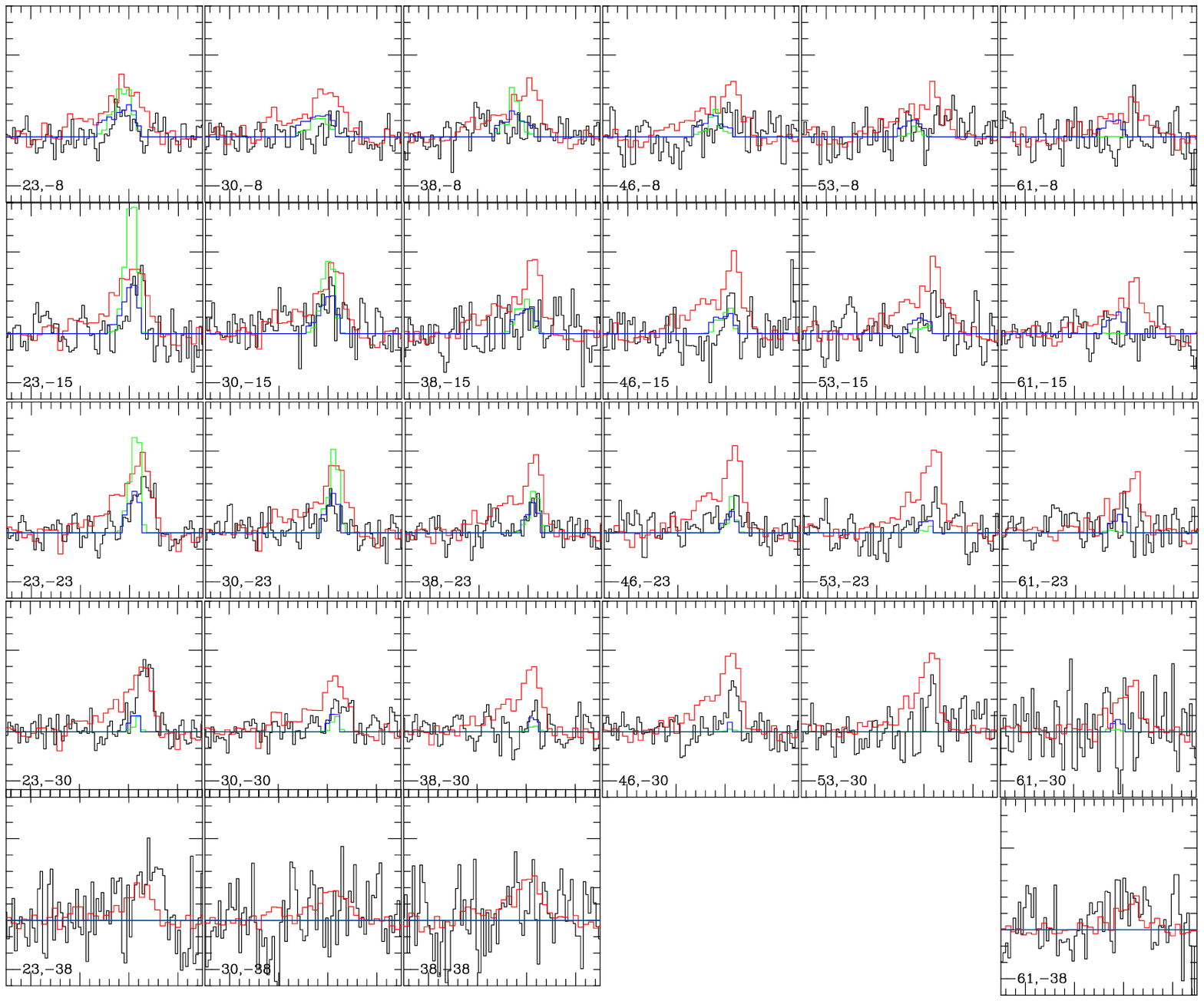}}
	\caption{IRAM 30m HERA $^{12}$CO(2-1) spectra of the southwestern box of Fig.~\ref{fig:pointingdss}
	  (black lines). H{\sc i} spectra from Kenney et al. (2004) (red lines). 
	  CO (green lines) and H{\sc i} (blue lines) spectra from the dynamical model of Vollmer et al. (2006).
	  The velocity scale is from 2150 to 2550~km\,s$^{-1}$ and the main-beam temperature scale 
	  from -40~mK to 80 mK (T$_{mb}$ scale).
	} \label{fig:cospec4}
\end{figure}

Fig.~\ref{fig:cospec1}-\ref{fig:cospec4} show the $^{12}$CO(2--1) HERA spectra (resolution of $11''$) 
together with the H{\sc i} spectra of Kenney et al. (2004) (resolution of $20''$) and the model
H{\sc i} and CO emission convolved to the observational resolutions (see Sect.~\ref{sec:comparison}).  
We observe that the H{\sc i} emission is more extended than the CO emission. Where detected, 
the velocity of the CO lines are close to the H{\sc i} velocities. 
In general, CO linewidths are comparable to the H{\sc i} linewidths in the galactic disk 
(Fig.~\ref{fig:cospec2}), but smaller in the extraplanar regions 
(Fig.~\ref{fig:cospec1}, \ref{fig:cospec4}).
A double-line profile is observed in the CO and H{\sc i} line west of the galaxy center 
at offsets $(-23,0),\ (-23,8)$
(Fig.~\ref{fig:cospec3}). This kind of line-profile has also
been observed in NGC~4438 (Vollmer et al. 2005) another Virgo spiral galaxy which undergoes ram pressure stripping
together with a tidal interaction.
In the southwestern part of the extraplanar gas the H{\sc i} profiles show a blueshifted wing
(Fig.~\ref{fig:cospec4}), corresponding to the most strongly pushed gas. Whereas the CO peak in this region is aligned with the H{\sc i} peak,
the blueshifted wing is absent in CO. This might be partly due to the smaller S/N ratio of the
CO data compared to the H{\sc i} data. Even if there is H$_2$ associated with the
blueshifted wing, we can conclude that the molecular fraction in this extraplanar
gas is lower in the blueshifted wing than in the main line.

\begin{figure}
	\resizebox{\hsize}{!}{\includegraphics{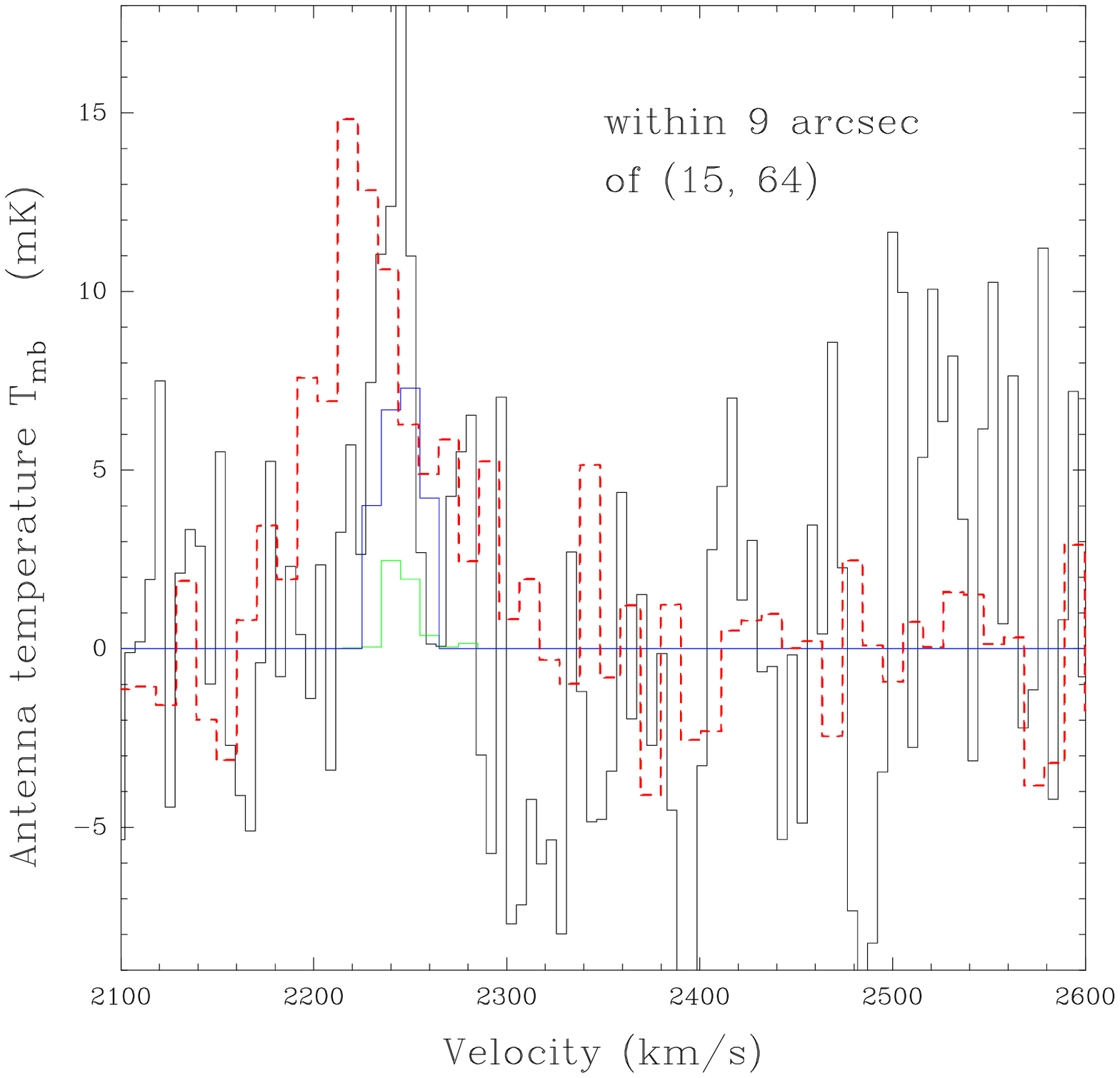}\includegraphics{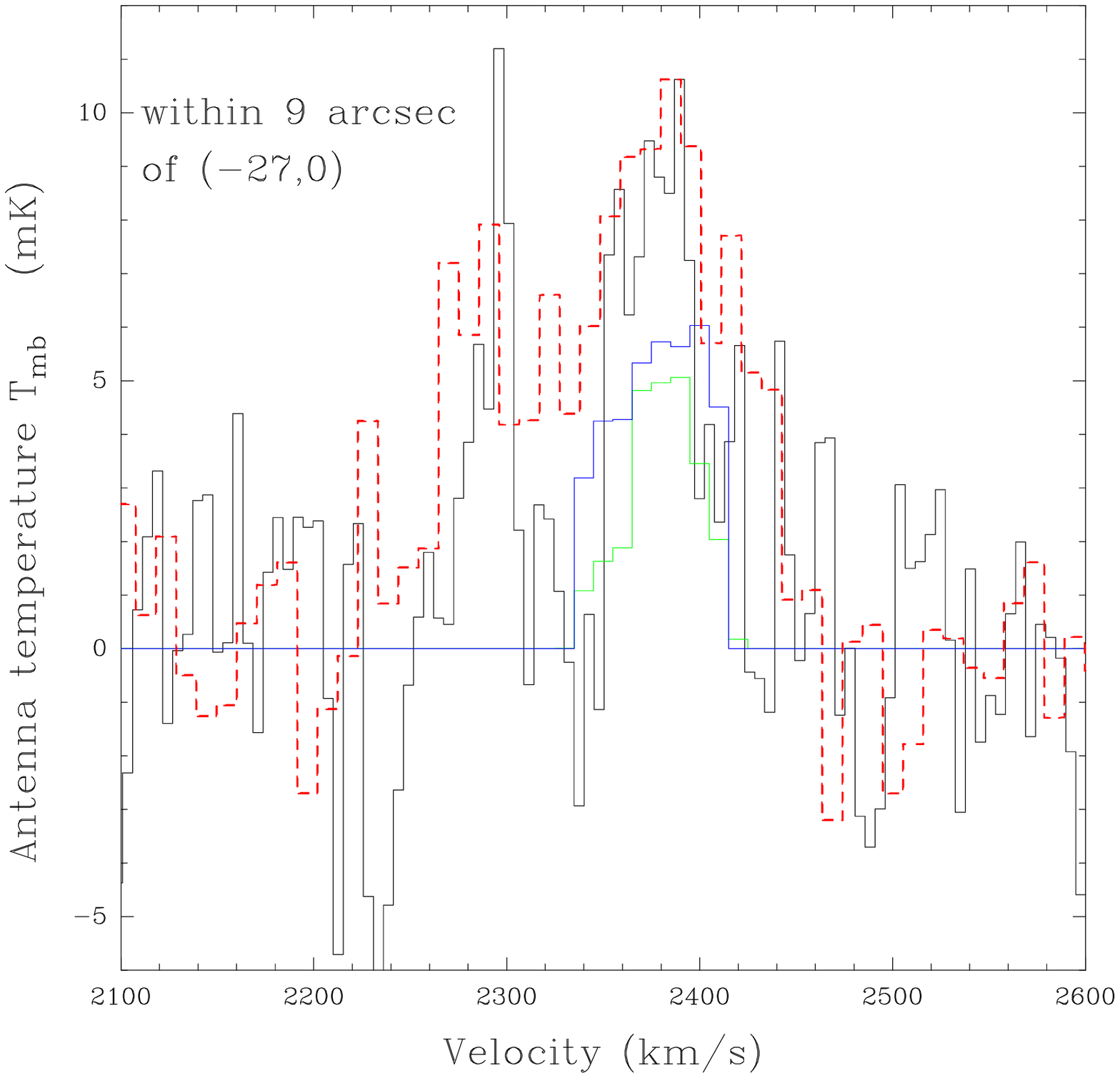}\includegraphics{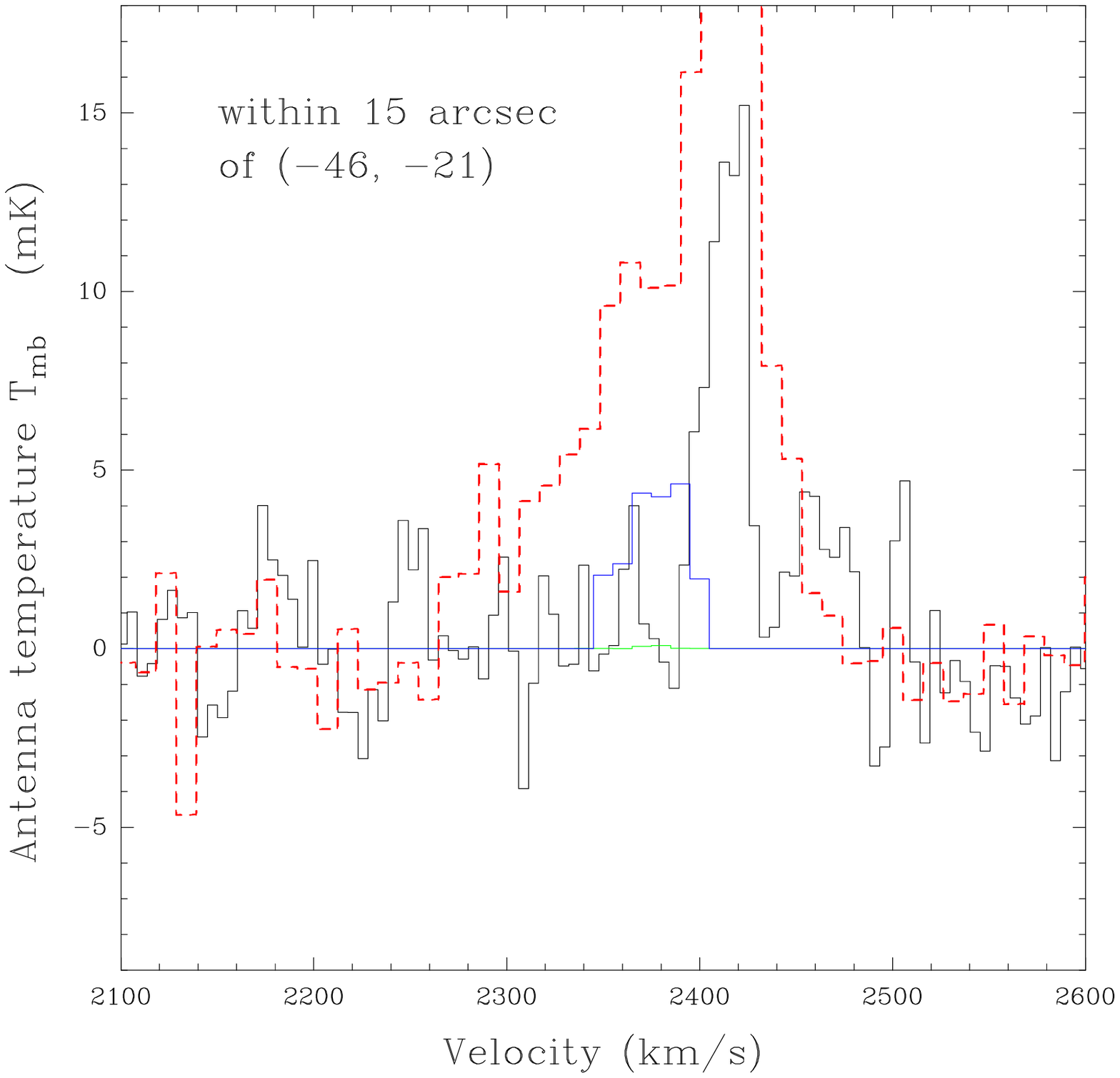}}
	\caption{Binned IRAM 30m HERA $^{12}$CO(2-1) spectra (see large circles on Fig.~\ref{fig:pointingdss}).
	  H{\sc i} spectra from Kenney et al. (2004) (red lines). 
	  CO (green lines) and H{\sc i} (blue lines) spectra from the dynamical model of Vollmer et al. (2006).
	} \label{fig:co2}
\end{figure}

Fig.~\ref{fig:co2} shows $^{12}$CO(2--1) HERA spectra of selected regions 
(large circles in Fig.~\ref{fig:pointingdss}, \ref{fig:pointinghi}). To obtain a better S/N ratio, 
the HERA spectra within each region were averaged.
These selected regions show the following characteristics:
\begin{itemize}
\item
CO emission is present in the northwestern extraplanar gas at offset $(15'',64'')$. 
The line is redshifted by $\sim 20$~km\,s$^{-1}$ with respect to the H{\sc i} line
(Fig.~\ref{fig:co2} left panel). 
\item
The double profile observed in the H{\sc i} line at offset $(-27'',0)$ is also present in the CO line.
The CO and H{\sc i} peak amplitudes and their separation are the same  (Fig.~\ref{fig:co2} middle panel).
\item
In the southwestern extraplanar gas at offset $(-46'',-21'')$ the blueshifted wing 
of the H{\sc i} line profile has no counterpart in the CO data (Fig.~\ref{fig:co2} right panel). 
The CO linewidth is significantly smaller
than the linewidth of the main H{\sc i} line. This confirms the low molecular fraction of the
blueshifted diffuse atomic gas. This blueshifted component corresponds to low surface density atomic 
gas discussed in Kenney et al. (2004; see their Fig.~10). The lower molecular fraction of this gas
confirms the claim of  Vollmer et al. (2006) that this gas has low densities making it more
vulnerable to ram pressure stripping.
\end{itemize}

\begin{figure}
        \psfig{file=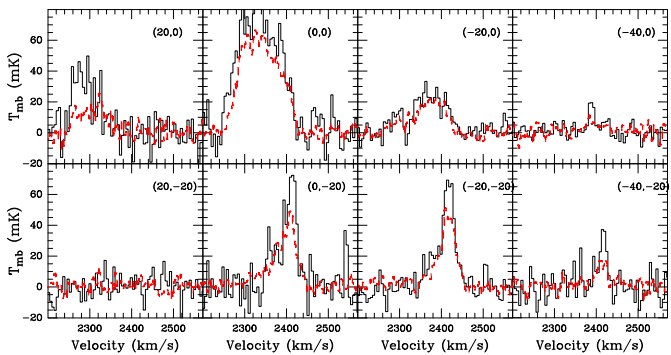,angle=-90,width=\hsize}
	\caption{Black: $^{12}$CO(1--0) spectra (see small circles on Fig.~\ref{fig:pointingdss}).
	  Red: convolved $^{12}$CO(2--1) HERA spectra.
	} \label{fig:co10}
\end{figure}

The $^{12}$CO(1--0) spectra (resolution: $21''$) are shown together with the convolved
$^{12}$CO(2--1) HERA spectra in Fig.~\ref{fig:co10}. 
For all but offset ($20'',0$), the CO(1--0) line closely follows the CO(2--1) line with a 
line ratio of $0.7 - 0.9$. Thus, the density and temperature of the molecular gas in the 
western extraplanar regions is probably not significantly different from those of the molecular gas
in the galactic disk.
At the eastern edge we observe a lower CO(2--1)/CO(1--0) ratio ($\sim 0.5$).
Given the sharpness of the gas distribution at the eastern edge and the
pointing uncertainty of the telescope at 115~GHz, this small line ratio
could be due to an offset of the CO(1--0) pointing to the west, i.e. closer to the major axis,
with respect to the CO(2--1) pointing.

\section{Wind-decoupled molecular gas \label{sec:decoupled}}

In another Virgo spiral galaxy, NGC~4438, Vollmer et al. (2005) found CO emission
not associated with any H{\sc i} emission. NGC~4438 underwent a tidal interaction
$\sim 100$~Myr ago (Combes et al. 1988) and now undergoes severe ram pressure stripping
(Vollmer et al. 2005). A narrow CO line was detected in the northern tidal arm 
of NGC~4438, with apparently no associated H{\sc i}.
Since the velocity of the CO line corresponds to that of the stellar component (determined using a dynamical model
of the tidal interaction), Vollmer et al. (2005) claimed that these molecular clouds
were too dense to be affected by ram pressure, i.e. that they decoupled
from the ram pressure wind.

\begin{figure}
	\psfig{file=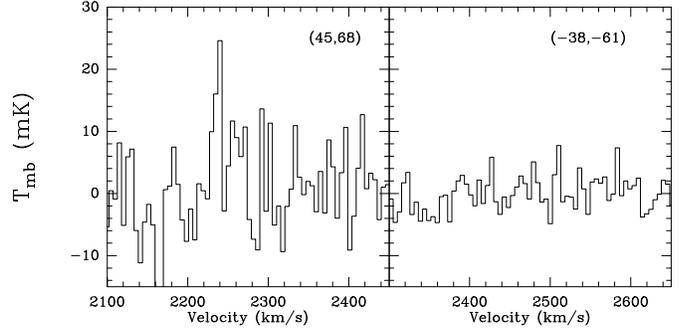,angle=-90,width=\hsize}
	\caption{Average $^{12}$CO(2-1) spectra of the northeastern $(45,68)$ and the
	  southwestern $(-38,-61)$ ends of the galactic disk, where the atomic hydrogen has been removed
	  by ram pressure. No H{\sc i} emission is detected nor does the model predict any gas.
	  The spectra are averaged over a region of $20''$.
	} \label{fig:diskspec}
\end{figure}

\begin{figure}
        \psfig{file=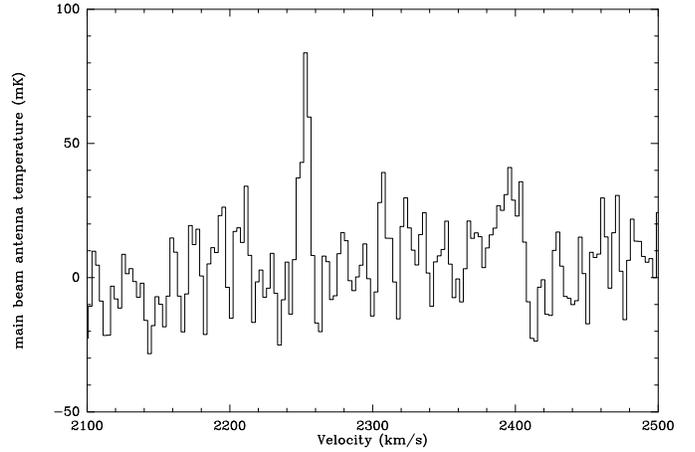,angle=-90,width=\hsize}
	\caption{Single $^{12}$CO(2-1) HERA spectrum in the northeast that is located outside
	  the H{\sc i} distribution. The position is marked with an additional box
	  in Fig.~\ref{fig:pointinghi}. 
	} \label{fig:decoupled}
\end{figure}

Based on these findings we searched for CO lines in regions devoid of any H{\sc i} emission
(i) in the ram pressure stripped outer galactic disk (Fig.~\ref{fig:diskspec}) and
(ii) between the northern galactic disk and the extraplanar H{\sc i} emission (Fig.~\ref{fig:decoupled}).
We only detect one CO line in the northern part of the galactic disk. The most prominent
CO line is detected at the position ($30,68)$; see the little box in Figs.~\ref{fig:pointingdss}
and \ref{fig:pointinghi}), i.e. between the stellar disk and the stripped extraplanar atomic gas.
The radial velocity of the CO lines is $\sim 2250$~km\,s$^{-1}$, close to the
velocity of the stellar component. The H{\sc i} line of the stripped
extraplanar H{\sc i} (Fig.~\ref{fig:cospec1}) is blueshifted by $\sim 30$~km\,s$^{-1}$.
We observe a clear absence of a CO line in the gas-free southwestern part of the disk
(Fig.~\ref{fig:diskspec} right panel), for which an explanation is proposed in Sect. 6.1.

\section{Comparison with the dynamical model \label{sec:comparison}}

In this section we will compare the molecular gas and H$\alpha$ distribution to the
dynamical model of Vollmer et al. (2006). 

\subsection{Molecular gas \label{sec:molcomp}}

\begin{figure}
	\resizebox{7cm}{!}{\includegraphics{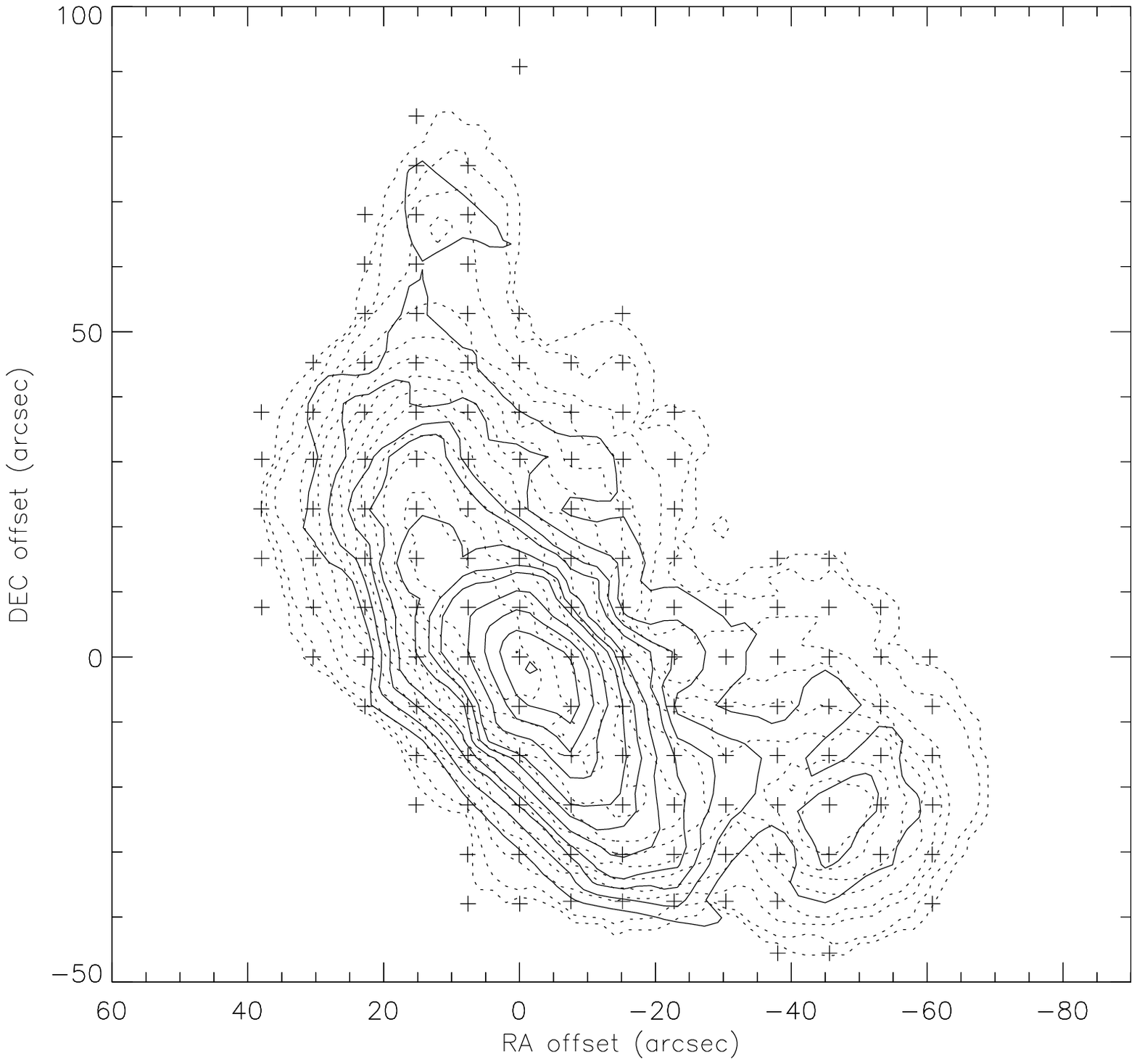}}
	\resizebox{7cm}{!}{\includegraphics{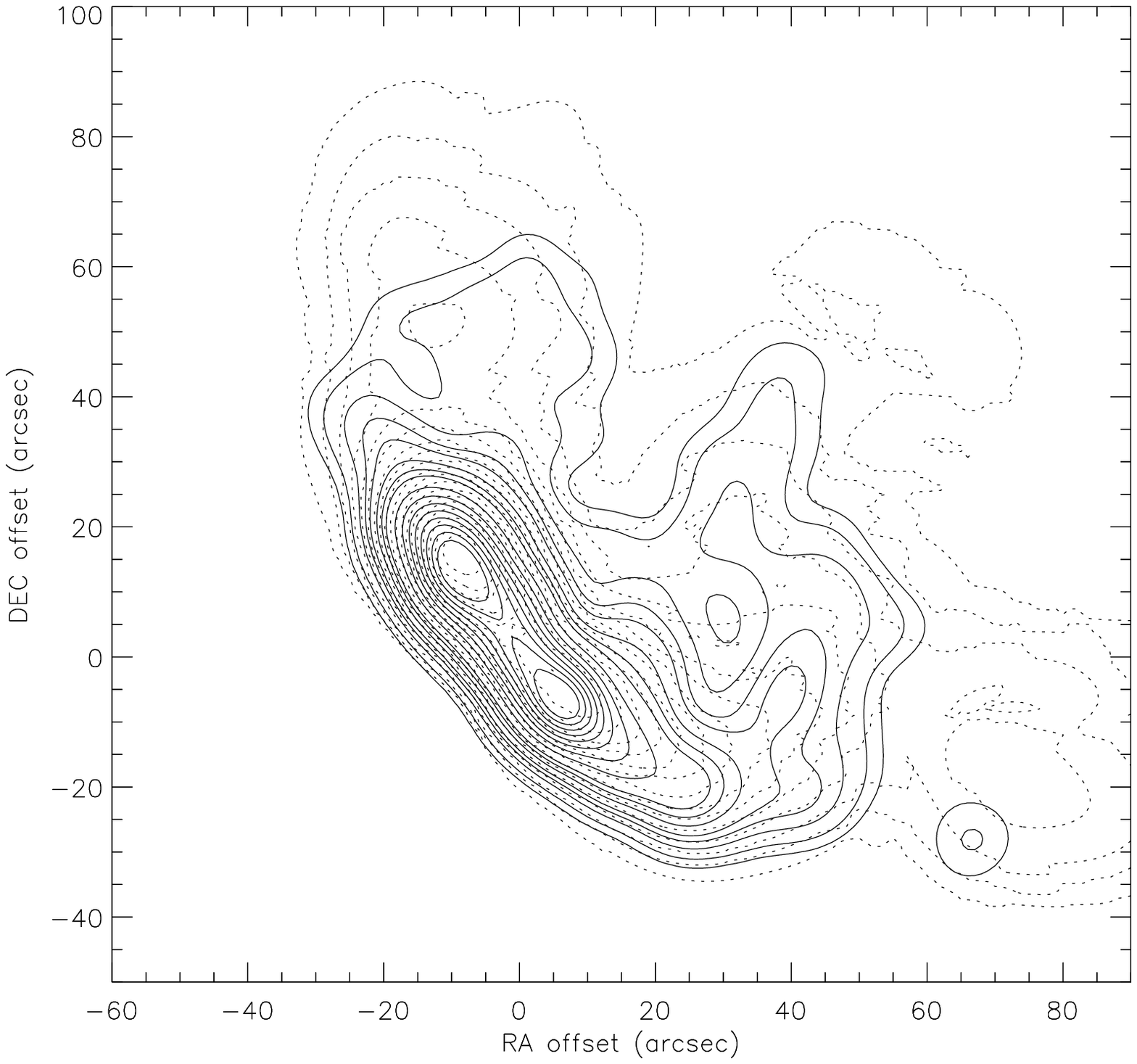}}
	\resizebox{7cm}{!}{\includegraphics{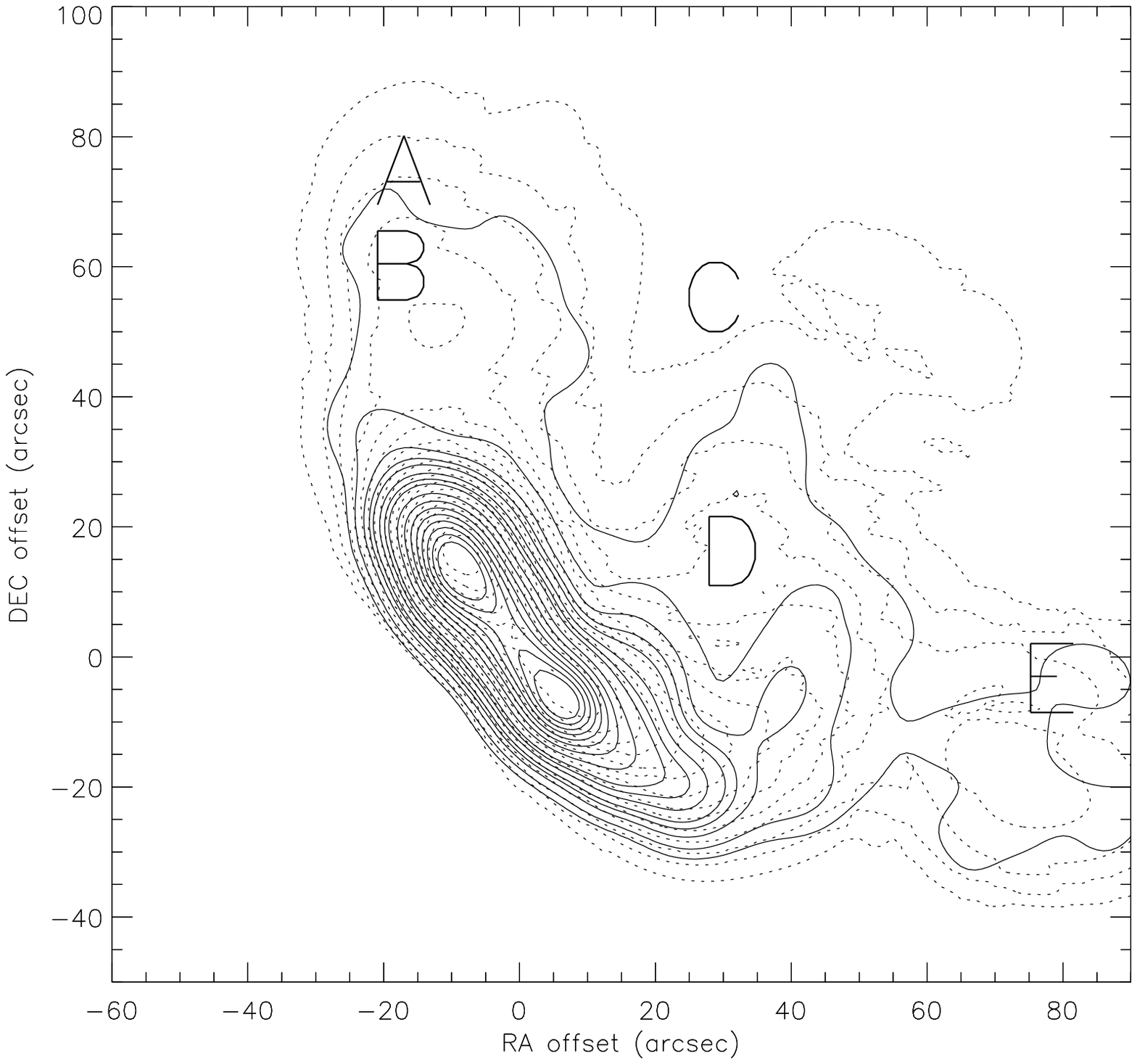}}
	\caption{Top panel: $^{12}$CO(2-1) emission distribution (solid line) on the
	  H{\sc i} emission distribution (dotted line). The plus signs show the IRAM HERA pointings
	  which were used to construct the CO moment 0 map. Middle panel: model CO
	  distribution (solid line) on the model H{\sc i} emission distribution (dotted line).
	  The gas is assumed to be entirely molecular for gas densities $\rho > 0.03$~M$_{\odot}$pc$^{-3}$. 
	  Lower panel: model CO
	  distribution (solid line) on the model H{\sc i} emission distribution (dotted line).
	  The molecular gas fraction is calculated using 
	  $f_{\rm mol}=\sqrt{\rho/(0.5\ {\rm M}_{\odot}{\rm pc}^{-3})}$.
	  The letters refer to the regions shown in the deprojection in Fig.~\ref{fig:deprojected}.
	} \label{fig:cohi}
\end{figure}

In order to estimate the H$_2$ column density distribution, via a zero 
moment (integrated intensity) map of the CO(2--1) emission, we proceed as below.
For the positions where the S/N of the H{\sc i} is higher than $2.5\sigma$,
the linewidths of the H{\sc i} spectra of these positions are determined and
the CO spectra integrated over the H{\sc i} velocity range. 
This leads to the CO emission distribution map of Fig.~\ref{fig:cohi}
(top panel).

To compare the H{\sc i} and the CO emission distributions with the dynamical model
of Vollmer et al. (2006), we assume that the molecular fraction of the gas depends on the
local gas density. In a first simple approach we assume that the gas is bimodal, i.e. entirely molecular
at densities $> 0.03$~M$_{\odot}$pc$^{-3}=1$~cm$^{-3}$ and entirely atomic at lower
gas densities. The resulting model gas distribution maps were convolved to the observational resolutions
(middle panel of Fig.~\ref{fig:cohi}).
In a second approach we assume that the molecular fraction depends linearly on the square root of the
gas density $f_{\rm mol}=\sqrt{\rho/(0.5\ {\rm M}_{\odot}{\rm pc}^{-3})}$. Moreover,
the molecular fraction cannot exceed unity. In Sec.~\ref{sec:molprop} we will give a motivation for this dependency.
The resulting model CO emission distribution is shown in the lower panel of Fig.~\ref{fig:cohi}. 

The comparison between the observed and the simulated CO emission distribution shows the following
similarities:
\begin{itemize}
\item
The galactic disk is the most prominent feature. There is more molecular gas of higher surface densities
in the outer part of the gas disk in the southwest than in the northeast.
\item
There is CO emission in the extraplanar regions.
\item
The model using a molecular fraction proportional to the square root of the gas density better reproduces
the CO morphology of the extraplanar regions: the northern and southwestern CO emission
regions are centered on the peaks of the H{\sc i} emission and there is a spatially
separated CO arm between the southwestern and the disk CO emission.
\end{itemize}
On the other hand, we observe the following disagreement between the model and our observations:
\begin{itemize}
\item
The model CO emission distribution shows a central hole. This is due to the initial
conditions which had an initial gas hole for computational reasons.
\item
The northern part of the observed CO disk shows emission of lower surface brightness.
The H{\sc i} emission (Kenney et al. 2004) shows the opposite trend: high 
column density gas is found to the north.
\item
The model CO emission to the west of the galaxy center is more extended than it is observed.
This is also the case for the H{\sc i} emission (see Sect.~\ref{sec:discussion}).
\end{itemize}
We thus conclude that the model reproduces qualitatively our CO observations.
The model using a density-dependent molecular fraction reproduces the observations better than 
a bimodal molecular fraction.

The direct comparison between the H{\sc i} and CO model and observed spectra 
(Fig.~\ref{fig:cospec1} to Fig.~\ref{fig:co2}) shows good agreement
for high intensities, i.e. high gas densities. However, the observed CO and H{\sc i} double lines and 
blueshifted wings of the H{\sc i} lines in the western extraplanar regions are not reproduced by the model.
This is due to the constant column density of the model gas in the regions affected by
ram pressure (see Sect.~\ref{sec:discussion}).

\subsection{Star formation}

In typical spiral galaxies the star formation rate follows more closely the molecular gas
distribution than the atomic gas distribution (Wong \& Blitz 2002).
The H{\sc i} surface density saturates at a value of $\sim10$~M$_{\odot}$pc$^{-2}$ or even declines for high
star formation rates per unit surface.
In an unperturbed galactic disk the ISM is confined in the gravitational potential of the disk.
The ISM is turbulent and this turbulence is most probably maintained by the energy input from SN explosions
(see, e.g. MacLow \& Klessen 2004 or Vollmer \& Beckert 2003).
Without a constant energy supply, turbulence is damped within a few Myr (Stone et al. 1998, MacLow 1999).
Since the extraplanar gas of NGC~4522 is no longer confined to the potential of the galactic disk,
it represents an ideal laboratory to test if the gravitational potential plays a role
for the correlation between star formation rate and the available molecular/atomic gas mass.

\begin{figure}
	\resizebox{\hsize}{!}{\includegraphics{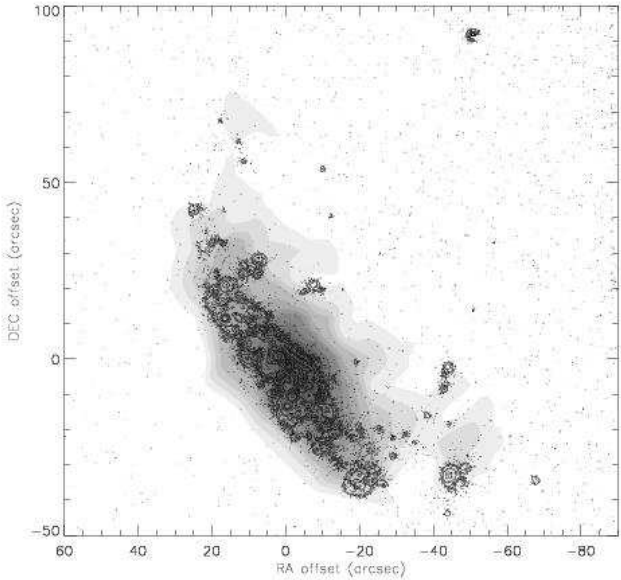}}
	\caption{H$\alpha$ emission distribution (Kenney et al. 2004; contours) on the 
	  $^{12}$CO(2-1) emission distribution (greyscale).
	} \label{fig:coha}
\end{figure}

To do so, we first present the H$\alpha$ emission distribution overlaid onto the CO emission
distribution (Fig.~\ref{fig:coha}).
There is no major site of massive star formation without associated CO emission.
At the edges and outside the galactic disk star formation does not coincide with the
maxima of CO emission. A high column density of molecular gas is thus not sufficient to
form massive stars. In other words, on kpc scales the star formation rate does not
directly depend on the molecular gas surface density.  A number of H{\sc ii} regions
are detected at the outer edges of the CO emission distribution, at offsets $(20'',45'')$, 
$(-10'',20'')$, $(-45'',0)$, $(-45'',-30'')$.

\begin{figure*}
	\resizebox{\hsize}{!}{\psfig{file=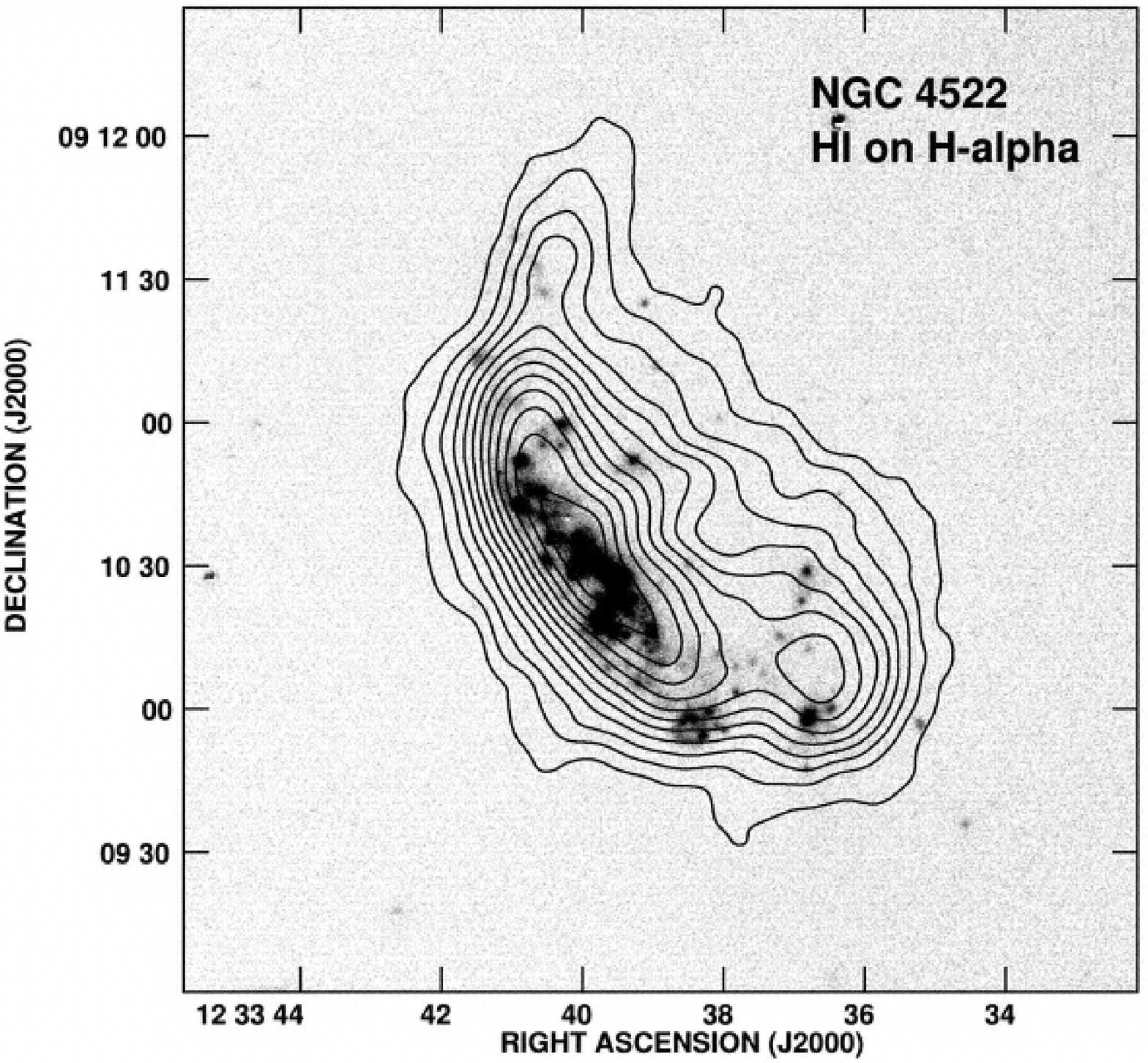,angle=-90,width=\hsize}\includegraphics{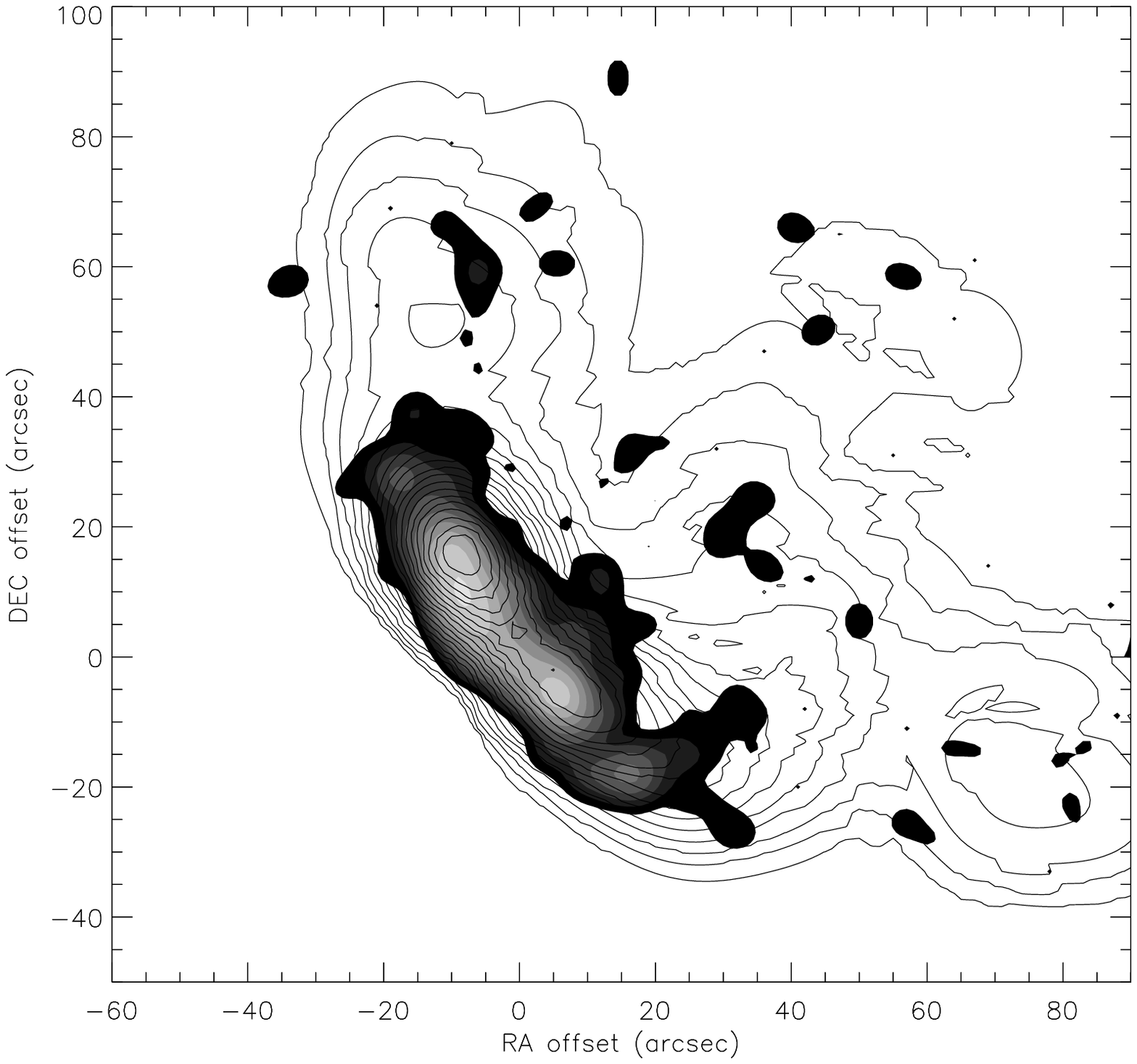}}
	\caption{Left panel: H{\sc i} emission distribution on H$\alpha$
	  emission distribution (Kenney et al. 2004). Right panel: model H{\sc i} distribution 
	  (Vollmer et al. 2006; contours) on the model massive star formation
	  distribution (greyscale). Darker regions correspond to less massive star formation.	  
	} \label{fig:coha2}
\end{figure*}

In the model we assume that the star formation rate is proportional to the number of collisions
between the gas clouds. Numerically, the star formation rate thus depends on the
local number density of the clouds, their cross section, and their local 3D velocity dispersion.
Since the model clouds have a constant surface density (Vollmer et al. 2006), their cross sections vary with
the cloud mass in the following way: $\pi r_{\rm cl}^2 = M_{\rm cl}/\Sigma$, where $r_{\rm cl}$ is
the radius, $M_{\rm cl}$ the cloud mass and $\Sigma$ the gas surface density.
The cloud mass distribution is a power law with an index of $-1.5$. 
For an isolated unperturbed spiral galaxy this prescription leads to a Schmidt law of the 
form $\dot{\Sigma}_{*} \propto \Sigma^{1.7}$, where $\dot{\Sigma}_{*}$ is the star formation rate
per area.
For the construction of a star formation distribution map we store all cloud--cloud
collisions during 20~Myr before the present state of the galaxy.
This is twice the timescale for H$\alpha$ emission, chosen to give more collisions and thus better statistics.
The distribution was then convolved to 0.6 times the resolution of the CO(2--1) map.
This model star formation distribution is presented in Fig.~\ref{fig:coha2} together with the 
H$\alpha$ emission distribution.
We note the following points for comparison between the model and observations:
\begin{itemize}
\item
As in the observations, our model has two prominent H{\sc ii} regions at the outer edges of the
galactic disk, the most prominent being the southwestern H{\sc ii} region.
These regions are separated by a local minimum from the rest of the galactic disk.
\item
We observe extraplanar model star formation regions close to the disk at the edges 
and in the middle of the galactic disk (see Sect.~\ref{sec:discussion}).
\item
There is isolated star formation over the whole extraplanar H{\sc i} emission.
In the model the small patches are due to multiple collisions of a single massive cloud.
\end{itemize}
We thus conclude that the model star formation distribution qualitatively reproduces
the observed H$\alpha$ emission distribution.

\section{Discussion \label{sec:discussion}}

\subsection{Deprojecting the model}

\begin{figure}
	\resizebox{\hsize}{!}{\includegraphics{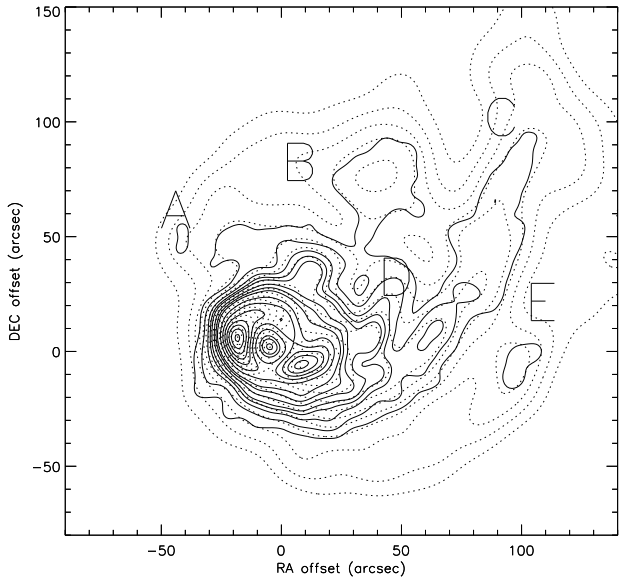}}
	\caption{Deprojected model CO distribution (solid line) on the deprojected H{\sc i} distribution
	  (dotted line). The letters correspond to the characteristic regions of Fig.~\ref{fig:cohi}.
	  The galaxy is moving towards the lower left corner and rotates clockwise.
	} \label{fig:deprojected}
\end{figure}

The model gives us the unique opportunity to deproject the gas distribution.
One has to keep in mind that the extraplanar gas is no longer located within the galactic disk,
but is a fully 3D feature.  In Fig.~\ref{fig:deprojected}, we present a deprojected face-on view of the 
model gas distribution of Fig.~\ref{fig:cohi} (lower panel). 
The corresponding regions are labeled with capital letters.
Region A is the tip of a spiral arm close to the main gas disk. Region B is a large overdensity within the 
stripped gas. Region C represents the tip of the most prominent gas (spiral) arm which is mainly made
of stripped material. Region D is a secondary gas arm and region E is an overdensity in the
windward, low-density side of the prominent gas (spiral) arm. As stated above, the gas
outside the main gas disk ($R<30''$) has a fully 3D structure.
As can be seen in the lower panel of Fig.~\ref{fig:cohi}, the upper edge of the deprojected
gas distribution becomes the northeastern edge of the model distribution in the observed projection.
The prominent gas (spiral) arm (D, C) runs in the observed projection vertically from the
southwestern edge of the galactic gas disk to the end of the gas distribution at $(40,40)$
(Fig.~\ref{fig:deprojected}). The southwestern extraplanar H{\sc i} emission region
is made of relatively diffuse gas from the low-density side of the prominent gas (spiral) arm.
This low-density side has one overdensity (E) where star formation can proceed.
This is consistent with the observed H{\sc i} and H$\alpha$ emission distribution, but we
do not claim that this is the necessary configuration.
Most interestingly, we can identify the upbending arm (south of D, 
C on the lower panel of Fig.~\ref{fig:cohi})
with a chain of observed H{\sc ii} regions (between ($12h33m38s$, $09d10m00s$) and
($12h33m37s$, $09d10m30s$) on the left panel of Fig.~\ref{fig:coha2} and the corresponding
molecular arm on Fig.~\ref{fig:coha}. 
We think that this is a robust feature.
As already stated in Sect.~\ref{sec:molcomp} the outer part of the upbending arm 
(DEC offset $0''$ to $40''$ in the lower panel of Fig.~\ref{fig:cohi}) is not present in the
observations. In deprojection this part of the arm is located between DEC offsets of
$20''$ and $80''$ (Fig.~\ref{fig:deprojected}). 
Due to its low total gas surface density $\Sigma$ with respect to the inner part (DEC offset $-20''$
to $0''$) it is more
vulnerable to ram pressure stripping or evaporation
(the acceleration $a$ due to ram pressure $p$ is $a=p/\Sigma$). 
This is consistent with the blueshift and large linewidth of the extraplanar low column density
H{\sc i} which is stripped more efficiently than the high column density gas (Fig.~9 of Vollmer et al. 2006).
Moreover, the CO and H{\sc i} spectra of the southern extraplanar H{\sc i} region (Fig.~\ref{fig:cospec4})
show that CO is only associated with high column density H{\sc i} at the highest velocities.
Our numerical model cannot reproduce the more efficient stripping of low column density gas,
because it uses a constant column density for the atomic gas phase (see Vollmer et al. 2001
and Fig.~9 of Vollmer et al. 2006).
We therefore suggest that the outer part of the stripped gas arm has been stripped more efficienctly
and now has a column density too low to be detected in the H{\sc i} observations.
On the other hand, very dense gas can decouple from the ram pressure wind (see Sect.~\ref{sec:decoupled})
as it is found in CO observations of NGC~4438 (Fig.~4 of Vollmer et al. 2005).

We conclude that the formation of molecular clouds and star formation mainly
depend on the large-scale overdensity of the gas more than on dynamical criteria
or the overall pressure (see Sect. 6.3).

Wind-decoupled molecular gas is only found in the northern part of the galactic disk
(Sec.~\ref{sec:decoupled}). With the help of the deprojection (Fig.~\ref{fig:deprojected}), 
this can be understood. High density gas is stripped from the left border of the gas disk.
Since the ram pressure wind is rather face-on, the gas clouds are pushed to 
larger heights above the galactic plane. At the same time, rotation makes clouds move towards
positions A and B, i.e. the northern part of the disk where the wind-decoupled molecular clouds
are found. This finding suggests that dense giant molecular clouds can decouple from the
ram pressure wind at early stages of the stripping of dense gas from the 
galactic disk, as seems to be the case for NGC~4438 as well.

\subsection{Molecular fraction \label{sec:molprop}}

As seen in Sec.~\ref{sec:molcomp} and Fig.~\ref{fig:cohi}, a molecular fraction which is
proportional to the square root of the total gas density reproduces the observed CO emission distribution
better than a simple bimodal molecular gas fraction assuming a gas density cutoff.
Vollmer \& Beckert (2003) approximated the molecular fraction by the ratio between the
turbulent crossing time scale $t_{\rm turb}=r_{\rm cl}/v_{\rm turb}$ and the
timescale for molecule formation $t_{\rm mol}=\alpha / \rho_{\rm cl}$, where $v_{\rm turb}$
is the turbulent velocity dispersion within the cloud, $\rho_{\rm cl}=\rho / \Phi_{\rm V}$ the
cloud density, $\rho$ the total large-scale gas density, $\Phi_{\rm V}$ the volume filling factor,
and $\alpha$ the constant of molecule formation $\alpha \sim 3 \times 10^7$~yr\,M$_{\odot}$pc$^{-3}$:
\begin{equation}
f_{\rm mol}=\frac{t_{\rm turb}}{t_{\rm mol}}=\frac{r_{\rm cl}\, \rho}{v_{\rm turb}\alpha \,\Phi_{\rm V}}\ .
\label{eq:mol1}
\end{equation}
The volume filling factor $\Phi_{\rm V}$ is defined by the condition that the gas clouds are self-gravitating, i.e.
the turbulent cloud crossing time $t_{\rm turb}$ equals the free fall time $t_{\rm ff}$ of the clouds:
\begin{equation}
\frac{r_{\rm cl}}{v_{\rm turb}}=\sqrt{\frac{3\,\pi\,\Phi_{\rm V}}{32\,G\,\rho}}\ ,
\label{eq:mol2}
\end{equation}
where $G$ is the gravitation constant.
Inserting Eq.~\ref{eq:mol2} into Eq.~\ref{eq:mol1} leads to
\begin{equation}
f_{\rm mol}=\sqrt{\frac{3\,\pi}{32\,G}} \alpha^{-1} \Phi_{\rm V}^{-\frac{1}{2}} \rho^{\frac{1}{2}}\ .
\label{eq:mol3}
\end{equation}
Thus, the molecular fraction depends on the square root of the cloud density $\rho_{\rm cl}=\Phi_{\rm V}^{-1}\rho$.
For the dependence used in Sec.~\ref{sec:molcomp} we assume a constant volume filling factor
(see Vollmer \& Beckert 2003). A density of $0.5$~M$_{\odot}$pc$^{-3}$ implies a
volume filling factor of $\Phi_{\rm V}=0.03$. This is higher than the volume filling factors
given in Vollmer \& Beckert (2003).  Eq.~\ref{eq:mol1} is
a crude approximation which may overestimate the molecular fraction by
a factor of 5-10, probably because $f_{mol}$ is substantially overestimated by $t_{turb}/t_{mol}$.  
However, we think that its dependence on the physical parameters of the gas are valid.
Eq.~\ref{eq:mol3} will  be used together with a similar expression for the star formation efficiency in
the next section.
The observed molecular gas fraction decreases from 50\,\% within the galactic disk to
35\,\% in the extraplanar region (Table~\ref{tab:table}).

\subsection{The efficiency of extraplanar star formation}

Does the star formation efficiency (SFE) change
once the gas has left the confining gravitational potential of the galactic disk?
The role of large-scale processes in provoking star formation is subject
to debate, with many "recipes" providing reasonably similar fits to observations --
e.g. the "Toomre" criterion (Kennicutt 1989), a pressure-based criterion (Blitz \& 
Rosolowsky 2006), or a basic Schmidt (1959) law.
The problem is not so much predicting the behavior in spiral disks but understanding 
what governs large-scale star formation in general in order to be able to understand other environments, typically
those at intermediate and high redshifts.  In classical dwarf galaxies, it is difficult to study the SFE because
the low metallicity makes the measure of the H$_2$ mass uncertain.  In Tidal Dwarf Galaxies,
morphologically similar but with higher metallicities, Braine et al (2001) found that, curiously, the
SFE was not identifiably different from spiral galaxies typically 100 times more massive.
In the post-collision Taffy galaxies or UGC~813/816 system, on the other hand, 
the SFE in the bridge gas is much lower than within spiral disks (Braine et al. 2003, 2004).

The SFE can be defined with respect to the molecular gas mass 
available or with respect to the total gas mass available, either
${\rm SFE}^{-1} = t_{\rm H_2}^{\rm SFR} = {\rm M(H_2)/\dot{M}_{SFR}}$  or
${\rm SFE}^{-1} = t_{\rm tot}^{\rm SFR} = {\rm M(H_2 + HI)/\dot{M}_{SFR}}$.
The observed gas masses, molecular fractions and star formation rates and timescales
are presented in Table~\ref{tab:table}.
\begin{table}
      \caption{Derived masses, molecular fractions, and star formation rates/timescales.}
         \label{tab:table}
      \[
       \begin{array}{|l|c|c|c|c|}
        \hline
         & {\rm galactic\ disk} & {\rm extrapl.} & {\rm total} & {\rm frac. extrapl.} \\
	\hline
	\hline
	M_{\rm HI}\ {\rm (10^8\ M_{\odot})} & 2.5 & 1.5 & 4 & 0.4 \\
	\hline
	M_{\rm H_2}\ {\rm (10^8\ M_{\odot})} & 2.2 & 0.8 & 3 & 0.25 \\
	\hline
	M_{\rm H_2}/M_{\rm HI} & 0.88 & 0.53 & 0.75 & \\
	\hline
	SFR\ {\rm (M_{\odot}yr^{-1})} & 0.1 & 0.015 & & \sim 0.14 \\
	\hline
	t_{\rm H_2}^{\rm SFR} {\rm (Gyr)} & 2.2 & 5.3 && \\
	\hline
	t_{\rm tot}^{\rm SFR} {\rm (Gyr)} & 4.7 & 15.3 && \\
	\hline
        \end{array}
      \]
\end{table}
Whereas close to half of the H{\sc i} is found beyond the galactic disk, this ratio
decreases to 1/4 for the molecular gas and to 1/7 for the H$\alpha$ emission.
The averaged star formation timescale based on the molecular or total gas mass increases from 
2.2 or 4.7~Gyr within the galactic disk to 5.3 or 15.3~Gyr respectively within the extraplanar region.
The star formation efficiency thus decreases by a factor of $\sim 3$ between the disk and the
extraplanar region.

In the framework of the model of Vollmer \& Beckert (2003) the local star formation rate
is given by
\begin{equation}
\dot{\rho}_* = \Phi_{\rm V} \frac{\rho}{t_{\rm ff}^{\rm cl}}\ ,
\label{eq:sfr}
\end{equation}
where $\Phi_{\rm V}$ is the probability of finding a self-gravitating cloud, i.e. the
volume filling factor of self-gravitating clouds.
Inserting the expression for the free fall time of Eq.~\ref{eq:mol2} into 
Eq.~\ref{eq:sfr} yields the following expression for the
star formation timescale which corresponds to the inverse of the star formation efficiency:
\begin{equation}
t_*=\frac{\rho}{\dot{\rho}_*}=\sqrt{\frac{3\,\pi}{32\,G}} \Phi_{\rm V}^{-\frac{1}{2}} \rho^{-\frac{1}{2}}\ .
\label{eq:sfr1}
\end{equation}
Thus, the star formation timescale depends on the inverse of the square root of the total large-scale
density and the volume filling factor.

From Table~\ref{tab:table} we obtain:
\begin{equation}
\frac{f_{\rm mol}^{\rm disk}}{f_{\rm mol}^{\rm ext}}=1.3\ {\rm ,\ and}\ \ \frac{t_*^{\rm ext}}{t_*^{\rm disk}}=3.3\ ,
\end{equation}
leading to
\begin{equation}
\frac{\Phi_{\rm V}^{\rm disk}}{\Phi_{\rm V}^{\rm ext}}=2.5\ {\rm ,\ and}\ \ \frac{\rho_{\rm disk}}{\rho_{\rm ext}}=4.3\ .
\label{eq:sfr2}
\end{equation}
We therefore suggest that the observed decrease of the star formation efficiency
by a factor of 3 in the extraplanar region, together with a lower molecular fraction,
is due to a higher volume filling factor of self-gravitating clouds in the galactic disk of NGC~4522.
These clouds are about twice as dense as their counterparts in the extraplanar regions.
The overall density in the extraplanar region is about 4 times lower than that of the galactic disk, presumably
because the extraplanar gas is no longer confined by the gravitational potential of the disk.
This is supported by the lower CO(2--1)/(1--0) line ratio in the extraplanar gas (Fig. 8).
However, the gas is still confined by the hot intracluster medium and partially compressed 
by ram pressure.  The mixture of ram-pressure and rotation create zones where the atomic gas is 
dense enough to be gravitationally bound, become molecular, and form stars.

We conclude that the stripped ISM still forms molecules and stars 
in a way not distinguishable from disk star formation (Eqs.~\ref{eq:mol1} and \ref{eq:sfr}) as long as the overall 
gas density is high enough to form bound clouds.  The ultimate fate of the stripped gas is 
probably ionization and evaporation, without star formation for the low-density gas, 
and after a generation of stars, which then disperse the remaining dense gas, for the initially denser gas.


As the simulations show (Fig.~\ref{fig:deprojected}) part of the gas is stripped in relatively dense arms
whose mean density is about 4 times lower than that
of the galactic gas disk. Since these gas arm are only confined by the hot intracluster medium,
they might ultimately disperse giving rise to a large low surface density tail as observed
in NGC~4388 (Oosterloo \& van Gorkom 2005).

\subsection{Comparison with radio continuum observations \label{sec:radio}}

\begin{figure*}
	\resizebox{\hsize}{!}{\includegraphics{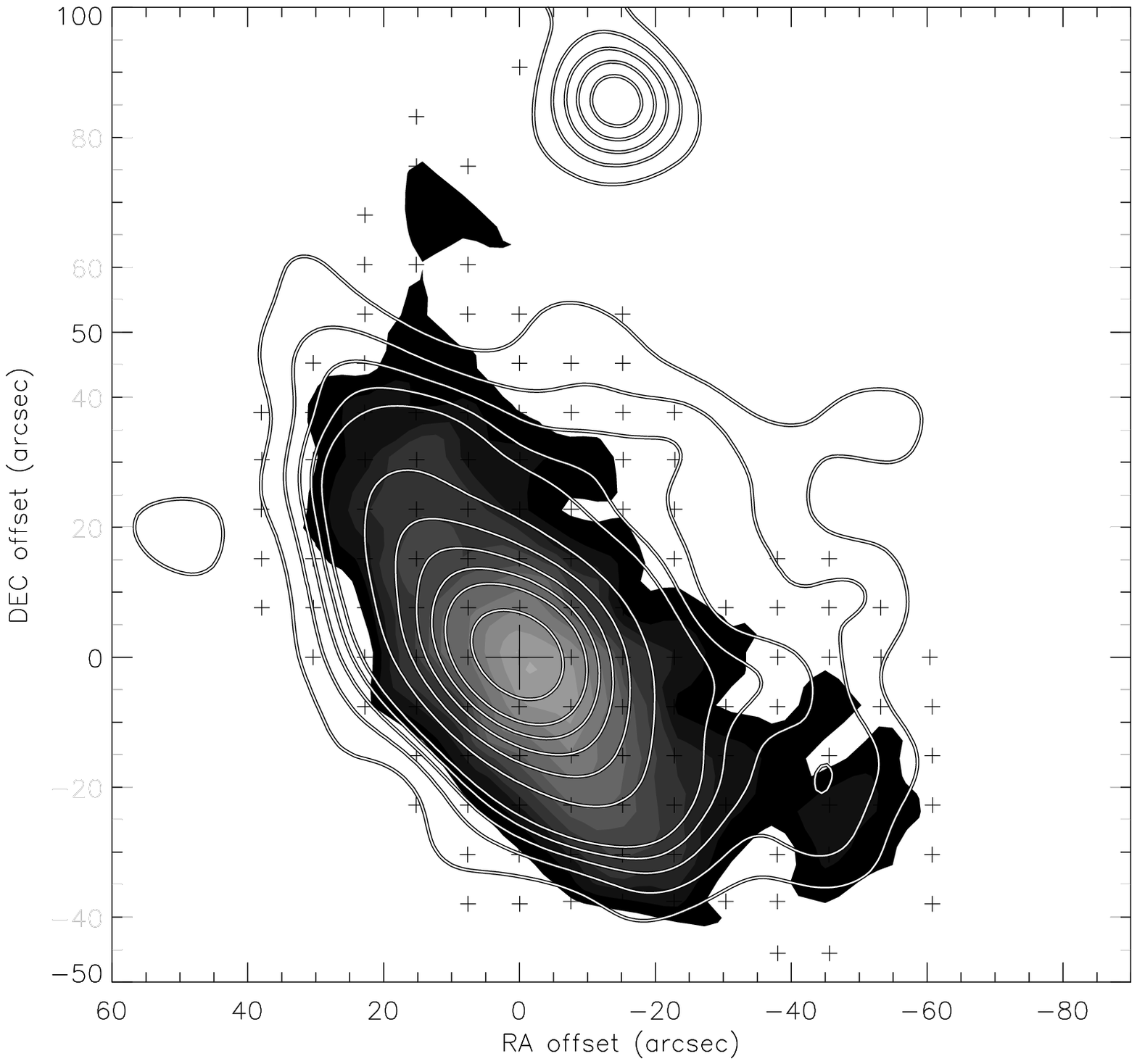}\includegraphics{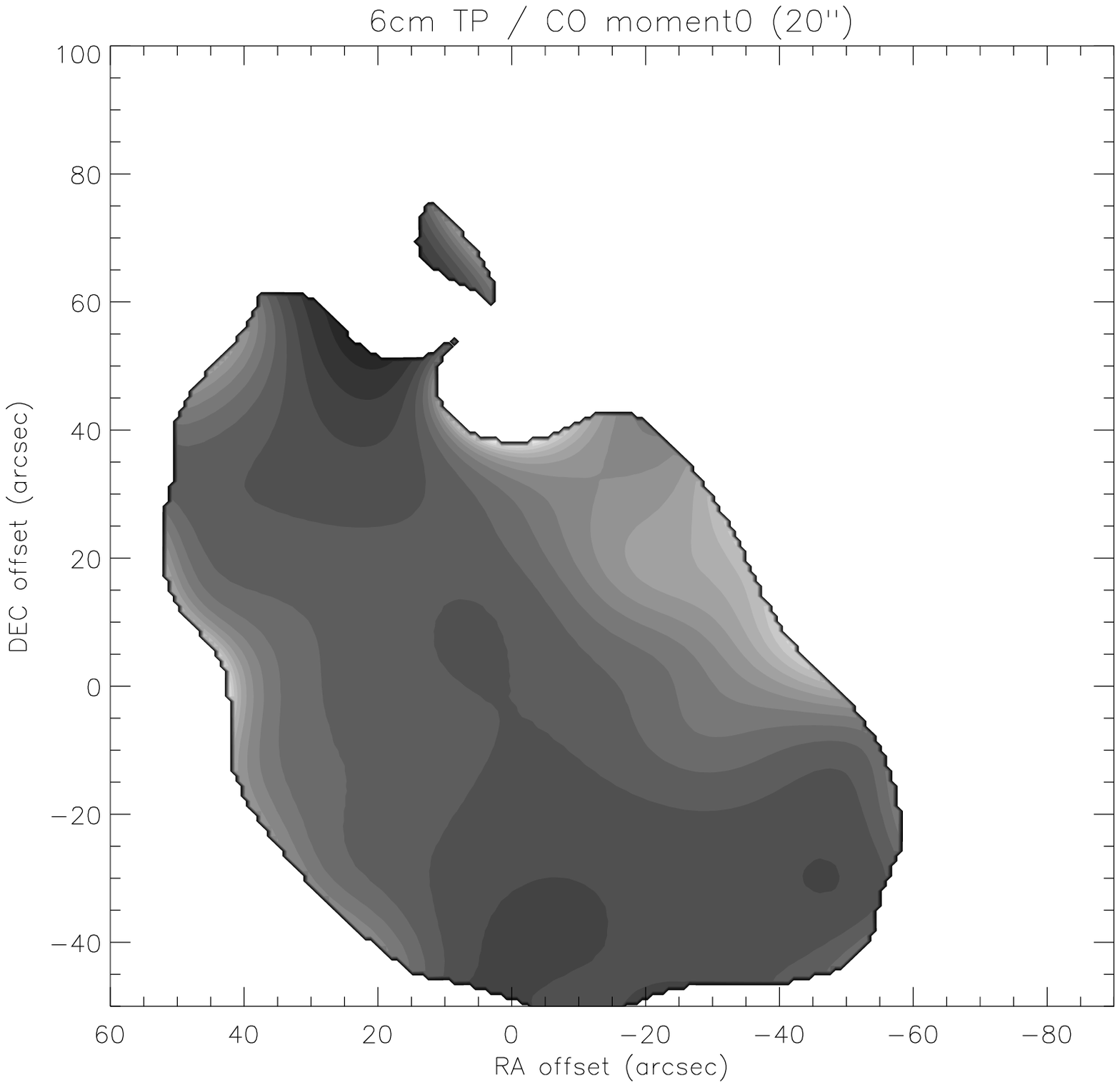}}
	\caption{Left panel: Contours of the 6~cm radio continuum emission from Vollmer et al. (2004)
	  on the CO emission distribution (greyscale). The contour levels are 
	  $(2,4,6,8,10,20,30,40,50,60,80,100) \times 20$~$\mu$Jy/beam. The resolution is $15'' \times 15''$.
	  Right panel: Ratio between the 6~cm and CO emission. Brighter regions have higher
	  CO/6~cm ratios.
	} \label{fig:co6cm}
\end{figure*}

Recently, Murphy et al. (2008a, 2008b) compared Spitzer $24$~$\mu$m emission with 20~cm
radio continuum maps. The radio-FIR correlation is used to predict the radio
emission from the Spitzer $24$~$\mu$m emission. They found a radio deficient region
at the eastern outer edge of NGC~4522's disk where ram pressure is pushing the
interstellar medium. Since the $24$~$\mu$m dust emission is associated with molecular gas,
we compare in Fig.~\ref{fig:co6cm} the 6~cm radio continuum emission from Vollmer et al.
(2004) with the CO emission distribution.
The extraplanar 6~cm radio continuum emission extends farther to the west than the CO emission.
The radio emission of the northeastern disk extends in the region where we found CO emission
without associated H{\sc i} emission (Fig.~\ref{fig:diskspec}). The northern extraplanar CO and H{\sc i}
emission does not show associated radio continuum emission. We convolved the 6~cm  
radio continuum and CO maps to a resolution $20'' \times 20''$ and computed a ratio map which is shown
in the right panel of Fig.~\ref{fig:diskspec}. As expected, the 6~cm/CO ratio is highest
in the extraplanar region. The smallest 6~cm/CO ratios are found at the extremities of the disk.
We observe a maximum of the 6~cm/CO at the eastern edge of the galactic disk where Murphy et al.
(2008a/b) detect a radio deficient region. Since we do not expect the $24$~$\mu$m emission to be
proportional to the CO emission, the interpretation of the result is difficult.
In addition, we think that our data are not sensitive enough to draw a firm conclusion on the
6~cm/CO ratio map.

\section{Conclusions \label{sec:conclusions}}

We present IRAM 30m $^{12}$CO(2--1) HERA and $^{12}$CO(1--0) observations of the ram pressure stripped
Virgo spiral galaxy NGC~4522. We directly compare the CO spectra to the H{\sc i} data cube
of Kenney et al. (2004). In a second step a CO emission distribution map is produced in the regions 
where H{\sc i} is detected. The CO emission distribution is compared to H$\alpha$ observations of
Kenney et al. (2004) and to the model distribution of molecular gas derived from dynamical
simulations of Vollmer et al. (2006). A map of the distribution of star formation based
on the numerical cloud--cloud collisions is produced which is then compared to the H$\alpha$ emission
distribution. The 3D model snapshot allows us to deproject the observed features
and understand their origin.
From this work we conclude that
\begin{enumerate}
\item
CO emission is associated with the extraplanar atomic gas. The morphology of the molecular gas closely follows 
but is less extended than the H{\sc i} morphology.
\item
In the northern part of the galactic disk we find CO emission without an H{\sc i} counterpart.
We interpret this detection as wind-decoupled molecular clouds as observed in NGC~4438 (Vollmer et al. 2005).
\item
In the extraplanar region CO emission is always associated with sites of massive star formation as 
probed by H$\alpha$ emission. At the resolution of our observations, there is no correlation between the
CO and H$\alpha$ peaks.
\item
A model using a molecular fraction proportional to the square root of the gas density qualitatively
reproduces our CO observations.
\item
The model star formation distribution, which is numerically based on cloud--cloud collisions, qualitatively
reproduces the observed H$\alpha$ emission distribution.
\item
The deprojection of the model extraplanar gas shows that a significant part of the gas is stripped in the form
of relatively dense arms.
\item
The formation of molecular clouds, and subsequent star formation, 
occurs at peaks in the large-scale volume density of the gas, with no 
clear difference with respect to the disk despite the very different conditions
(i.e. stellar density dominates in the disk but is negligible in the extraplanar material).
\item
In the disk gas, the molecular and atomic fractions are about equal 
whereas in the extraplanar gas, there is twice as much HI as H$_2$, assuming 
a standard $N({\rm H}_2) / I_{\rm CO}$ conversion ratio.
\item
The star formation efficiency of the extraplanar gas is about 3 times lower than that of the
galactic disk.
\item
Using the analytical framework of Vollmer \& Beckert (2003) we find that the overall total gas density and
volume filling factor of self-gravitating clouds are factors of 4 and 2.5 lower respectively compared 
to the galactic disk. 
\end{enumerate}
In the early phases of ram pressure stripping ($\sim 50$~Myr after peak ram pressure; Vollmer et al. 2006)
a significant part of the stripped gas is in the form of relatively dense arms. 
At the same time some very dense molecular clouds, representing a tiny fraction of the
stripped gas, can decouple from the ram pressure wind.
Molecules and stars form
within the stripped dense gas according to the same laws as in the galactic disk, i.e. they mainly depend on the
overall total gas density. Star formation proceeds where the local large-scale gas density is highest. 
Given the complex 3D morphology this does not necessarily correspond to the peaks of the surface density.
In the absence of a confining gravitational potential these stripped gas
arms will most probably disperse, i.e. their density will decrease and star formation will cease.

\begin{acknowledgements}
Based on IRAM observations. IRAM is supported by INSU/CNRS (France), MPG (Germany), and IGN (Spain). 
We made use of a DSS image. The Digitized Sky Survey was produced at the Space Telescope Science Institute 
under U.S. Government grant NAG W-2166. The images of these surveys are based on photographic data obtained 
using the Oschin Schmidt Telescope on Palomar Mountain and the UK Schmidt Telescope. The plates were
processed into the present compressed digital form with the permission of these institutions.
\end{acknowledgements}

\end{document}